\documentclass[aps,prb,twocolumn,showpacs]{revtex4}

\usepackage{bbm}
\usepackage{amssymb,amsmath}
\usepackage{graphicx}
\begin{document}
\title{Tight-Binding model for semiconductor nanostructures}
\author{S. Schulz}
\affiliation{Institute for Theoretical Physics, University of Bremen, D-28334
Bremen, Germany}
\author{G. Czycholl}
\affiliation{Institute for Theoretical Physics, University of Bremen, D-28334
Bremen, Germany}
\date{\today}
\begin{abstract}
An empirical $s_cp^3_a$ tight-binding (TB) model is applied to the
investigation of electronic states in  semiconductor quantum dots. A
basis set of three $p$-orbitals at the anions and one $s$-orbital at
the cations is chosen. Matrix elements up to the second nearest
neighbors and the spin-orbit coupling are included in our TB-model.
The parametrization is chosen so that the effective masses, the
spin-orbit-splitting and the gap energy of the bulk CdSe and ZnSe
are reproduced. Within this reduced $s_cp_a^3$ TB-basis the valence
(p-) bands are excellently reproduced and the conduction (s-) band
is well reproduced close to the $\Gamma$-point, i.e. near to the
band gap. In terms of this model much larger systems can be
described than within a (more realistic) $sp^3s^*$-basis. The
quantum dot is modelled by using the (bulk) TB-parameters for the
particular material at those sites occupied by atoms of this
material. Within this TB-model we study pyramidal-shaped CdSe
quantum dots embedded in a ZnSe matrix and free spherical CdSe
quantum dots (nanocrystals). Strain-effects are included  by using
an appropriate model strain field. Within the TB-model, the
strain-effects can be artifically switched off to investigate the
infuence of strain on the bound electronic states and, in
particular, their spatial orientation. The theoretical results for
spherical nanocrystals are compared with data from tunneling
spectroscopy and optical experiments. Furthermore the influence of
the spin-orbit
coupling is investigated.%
\end{abstract}

%
\maketitle
\section{Introduction}
\label{intro} Semiconductor quantum dots~\cite{woggon97,michler2000}
(QDs) are of particular interest, both concerning basic research and
possible applications. QDs are considered to be zero dimensional
objects, i.e. systems confined in all three directions of space with
a typical size of the magnitude of several nanometers. Therefore,
these systems are realizations of ``artificial atoms'' whose form
and size can be manipulated. Concerning basic research these
nanostructures (QDs) are interesting, as the methods of quantum
theory can be applied to systems on new scales and with new
symmetries in between that of atoms or molecules and of macroscopic
crystals. On the other side light emission and absorption just from
the localized states in such devices may be important for
optoelectronic applications~\cite{passaseo2001,grundmannnano},
quantum cryptography~\cite{michlerscience2000} and quantum
computing~\cite{li2003}.

Semiconductor QDs can be realized by means of metallic gates
providing external (electrostatic) confinement
potentials~\cite{kouwenhoven2003}, by means of selforganized
clustering of certain atoms in the Stranski-Krastanow (SK) growth
mode~\cite{merz98,yiyang2001,zhang2001} or chemically by stopping
the crystallographic growth using suitable surfactant
materials~\cite{kim2001,guzelian96,manna2000,murray2001}. Here we
deal only with the latter two types of QDs. The QDs created in the
SK growth mode emerge self-assembled or self-organized in the
epitaxial growth process because of the preferential deposition of
material in regions of intrinsic strain or along certain
crystallographic directions. In epitaxial growth of a semiconductor
material A on top of a semiconductor material B only one or a few
monolayers of A material may be deposited homogeneously  as a quasi
two dimensional (2d) A-layer on top of the B-surface forming the so
called wetting layer (WL). Under certain conditions and for certain
materials further deposited A-atoms will not form a further
homogeneous layer but they will cluster and form islands of
A-material because this may lower the elastic energy due to the
lattice mismatch of the A- and B-material. If one then stops the
growth process, one has free A-QDs on top of an A-WL on the
B-material. If one continues the epitaxial growth process with
B-material, one obtains embedded quantum dots (EQDs), i.e. QDs of
A-material on top of an A-WL embedded within B-material.\\
The chemically realized QDs emerge
 by means of colloidal chemical
synthesis~\cite{kim2001,guzelian96}. Thereby the crystal growth of
semiconductor material in the surrounding of soap-like films called
surfactants is stopped when the surface is covered by a monolayer of
surfactant material. Thus one obtains tiny crystallites of the
nanometer size in all three directions of space, why these QDs are
also called ``nanocrystals'' (NCs). The size and the shape of the
grown NCs can be controlled by external parameters like growth time,
temperature, concentration and surfactant
material~\cite{manna2000,murray2001}. Certain physical properties
like the band gap (and thus the color) depend crucially on the size
of the NCs. Typical diameters for both, EQDs and NCs, are between 3
and 30 nm, i.e. they contain between $10^3$ up to $10^5$ atoms.
Therefore, EQDs and NCs can be considered to be a new, artficial
kind of condensed matter in between molecules and solids. For the in
SK-modus grown EQDs lens-shaped dots~\cite{wojs96}, dome shaped and
pyramidal dots~\cite{fry2000,ruvimov95,merz98}, and also truncated
cones~\cite{shumway2001} have been found and considered.

Of course the fundamental task is the calculation of the electronic
properties of EQDs and NCs. But here one encounters the difficulty
that these systems are much larger than conventional molecules and
that the fundamental symmetry of solid state physics, namely
translational invariance, is not fulfilled. Therefore, neither the
standard methods of theoretical chemistry nor the ones of solid
state theory can immediately be applied to systems with up to $10^5$
atoms. Conventional ab-initio methods of solid state theory based on
density functional theory (DFT) and local density approximation
(LDA) would require supercell calculations. But the size of a
supercell must be larger than the EQD or NC, and such large
supercells are still beyond the possibility of present day
computational equipment. Therefore, only systems with up to a few
hundred atoms can be investigated in the framework of the standard
ab-initio DFT methods~\cite{puzder2004,sarkar2003,deglmann2002}.
Simple model studies based on the effective mass
approximation~\cite{wojs96,grundmann1} or a multi-band
$\vec{k}\cdot\vec{p}$-model~\cite{grundmann98,fonoberov2003,cpryor98}
describe the QD by a confinement potential caused by the band
offsets, for instance; they give qualitative insights into the
formation of bound (hole and electron) states, but they are too
crude for quantitative, material specific results or predictions.
More suitable for a microscopic description  are empirical
pseudopotential methods~\cite{wang99,wang2000,williamson2000,kim98}
as well as empirical tight-binding models
~\cite{saito2002,lee2001,lee2002,allan2000,niquet2001,santo,
hill94,leung98,leung99,pokrant99,schrier2003,lee2004}. The empirical
pseudopotential methods allow for a detailed variation of the wave
functions on the atomic scale. This is certainly the most accurate
description from a microscopic, atomistic viewpoint, but it requires
 a large set of basis states. Within a TB-model some kind of coarse graining
is made and one studies spatial variations only on inter-atomic scales and
no longer within one unit cell. The advantage  is that usually a small basis set
is sufficient, which allows for the possibility to study larger systems.
Furthermore the TB-model provides a simple
physical picture in terms of the atomic orbitals and on-site and inter-site
matrix elements between these orbitals. A cutoff after a few
neighbor shells is usually justified for orbitals localized at the
atomic sites.

Semiempirical TB-models have been used already to describe
''nearly'' spherical InAs and CdSe NCs for which the dangling bonds
at the surfaces are saturated by
hydrogen~\cite{lee2001,lee2002,niquet2001,allan2000} or organic
ligands~\cite{leung98,leung99,pokrant99}. Also
uncapped~\cite{saito98} and capped~\cite{santo} pyramidal InAs QDs
were ivestigated by use of an empirical TB-model. In the latter work
an $sp^3s^*$-basis was used leading to a $10N \times 10N$
Hamiltonian matrix, where $N$ is the number of atoms, with 33
independent parameters. In the present paper we apply a similar
TB-model to II-VI nanostructures, namely CdSe EQDs embedded within
ZnSe and spherical CdSe NCs. We show that a smaller TB-basis is
sufficient, namely an $s_cp_a^3$-basis, i.e. 4 states per unit cell
and spin direction. This requires only 8 independent parameters and,
in principle, allows for the investigation of larger nanostructures
than were accessible in Ref.~\onlinecite{santo}. Strictly speaking,
the $s_cp^3_a$-basis-set leads to a smaller matrix-dimension and
also to a smaller number of nonzero matrix elements compared to
$sp^3s^*$ TB-model. So the $s_cp^3_a$ TB-model is numerically less
demanding regarding both memory requirements and computational time.
For the bulk system the valence p-bands are excellently reproduced
and the conduction s-band is well reproduced close to the
$\Gamma$-point. Therefore, we expect that also for the QDs all the
hole states and at least the lowest lying electron states (close to
the gap) are well reproduced. We investigate, in particular, the
influence of strain effects on the electronic structure. To examine
the accuracy of our model we compare the results to other
microscopic and macroscopic models. Furthermore TB-results obtained
for CdSe-NCs are compared to experimental results, and very good
agreement, for instance  for the dependence of the energy gap on the
NC-diameter, is obtained. This demonstrates that our TB-model with a
reduced basis set is reliable and sufficient for the reproduction of
the most essential electronic properties of the nanostructures.
\begin{table}
\begin{center}

\begin{tabular}{ccc}
\hline\hline\noalign{\smallskip}
 & CdSe & ZnSe  \\
\noalign{\smallskip}\hline\noalign{\smallskip}
$E_{\text{g}}$ [eV] & 1.74~\cite{kim94} & 2.8201~\cite{kim94} \\
$\Delta_{\text{so}}$ [eV] & 0.41~\cite{kim94} & 0.43~\cite{kim94} \\
$m_{\text{e}}$ & 0.12~\cite{kim94} & 0.147~\cite{hoelscher85} \\
$\gamma_1$ & 3.33~\cite{kim94} & 2.45~\cite{hoelscher85} \\
$\gamma_2$ & 1.11~\cite{kim94} & 0.61~\cite{hoelscher85} \\
$\gamma_3$ & 1.45~\cite{kim94} & 1.11~\cite{hoelscher85} \\
$C_{12}$ [GPa] & 46.3~\cite{pelle95} & 50.6~\cite{pelle95} \\
$C_{11}$ [GPa] & 66.7~\cite{pelle95} & 85.9~\cite{pelle95} \\
\noalign{\smallskip}\hline\hline
\end{tabular}
\caption{Properties of the CdSe and ZnSe bandstructures. The lattice
constants are given by 6.077 \AA\ and 5.668 \AA, respectively.
$E_{\text{g}}$ denotes the band gap, $\Delta_{\text{so}}$ the
spin-orbit coupling and $m_{\text{e}}$ the effective electron mass.
The Kohn-Luttinger-Parameters are $\gamma_1$,$\gamma_2$ and
$\gamma_3$. The $C_{ij}$ are the elements of the elastic stiffness
tensor.} \label{tab:materialpara}
\end{center}
\end{table}

This work is organized as follows. In Sec.~\ref{sec:formalism} our
TB-model is presented. The formalism how to obtain the TB-parameters
and how to apply them  to the description of EQDs and NCs is
described. In Sec.~\ref{sec:resultspyramida} the inclusion of strain
effects in our model is introduced. Results for the pyramidal CdSe
EQDs are presented. For the spherical CdSe NCs the results and the
comparison with the experimental data are presented in
Sec.~\ref{sec:resultsnano}. Section~\ref{sec:conclusion} contains a
summary and a conclusion.

\section{Theory}
\label{sec:formalism}
\subsection{TB-Model for bulk materials}
\label{sec:2a}
In this work we use a TB-Model with 8 basis states per unit cell.
Such a model has been succesfully used for the
investigation of optical properties in ZnSe-quantum wells~\cite{dierkspaper98}.
For the description of the bulk semiconductor compounds CdSe and ZnSe we
choose an $s^{}_cp^3_a$ basis set. That implies that the set
of basis states $|\nu,\alpha,\sigma,\vec{R}\rangle$ is given by four orbitals $\alpha=s,p_x,p_y,p_z$ with
spin $\sigma=\pm\frac{1}{2}$. One $s$-orbital at the cation ($\nu=c$) and three
$p$-orbitals at the
anion ($\nu=a$) site in each unit cell $\vec{R}$ are chosen. The TB matrix elements are given by
\begin{equation}
E_{\alpha,\alpha'}(\vec{R}'-\vec{R})_{^{\nu,\nu'}}=
\langle\nu',\alpha',\sigma'\vec{R}'|H^{\text{bulk}}|\nu,\alpha,\sigma,\vec{R}\rangle \; .
\end{equation}

\begin{figure}
\begin{center}
\resizebox{0.49\textwidth}{!}{%
  \includegraphics{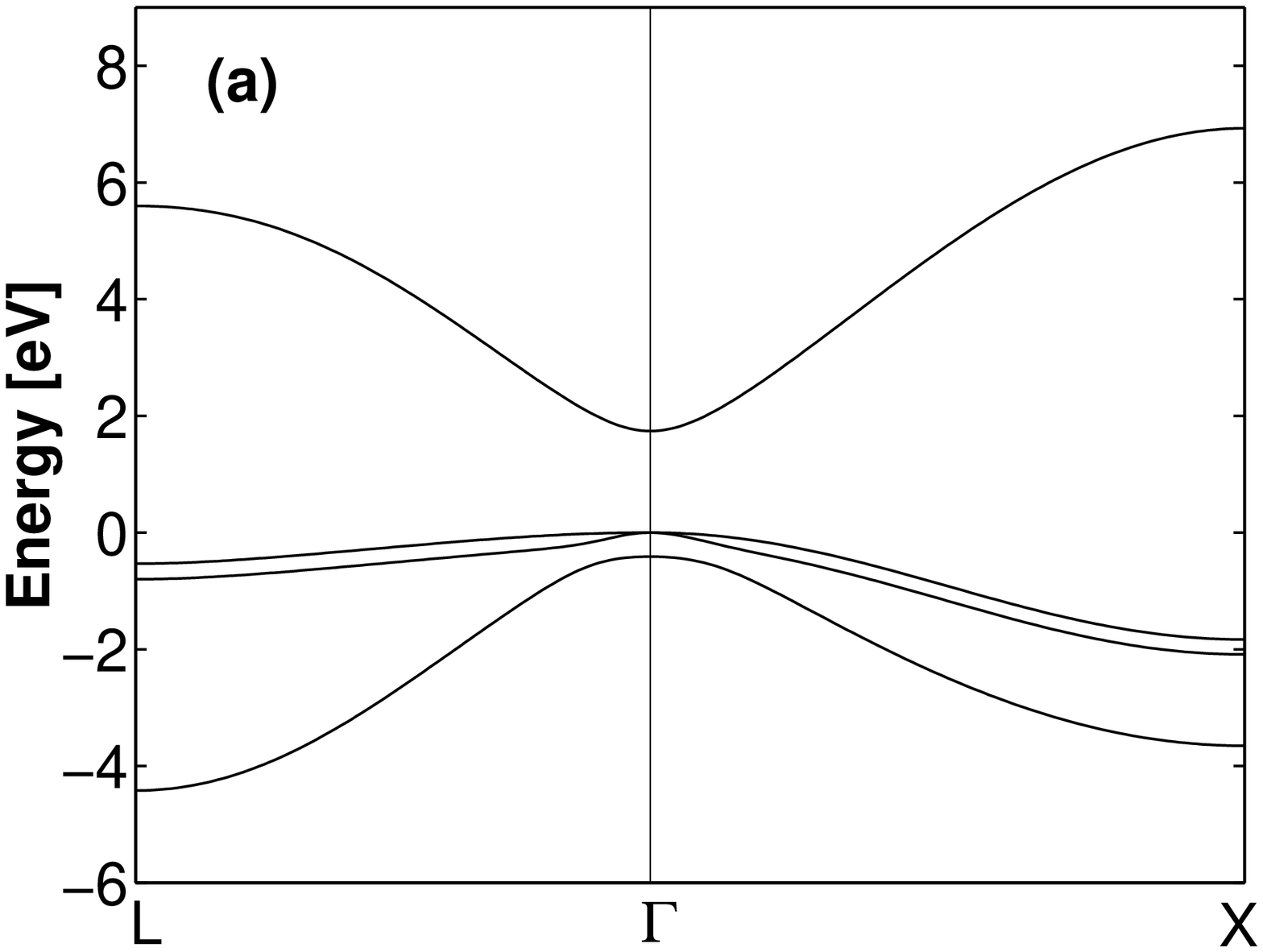}
  \includegraphics{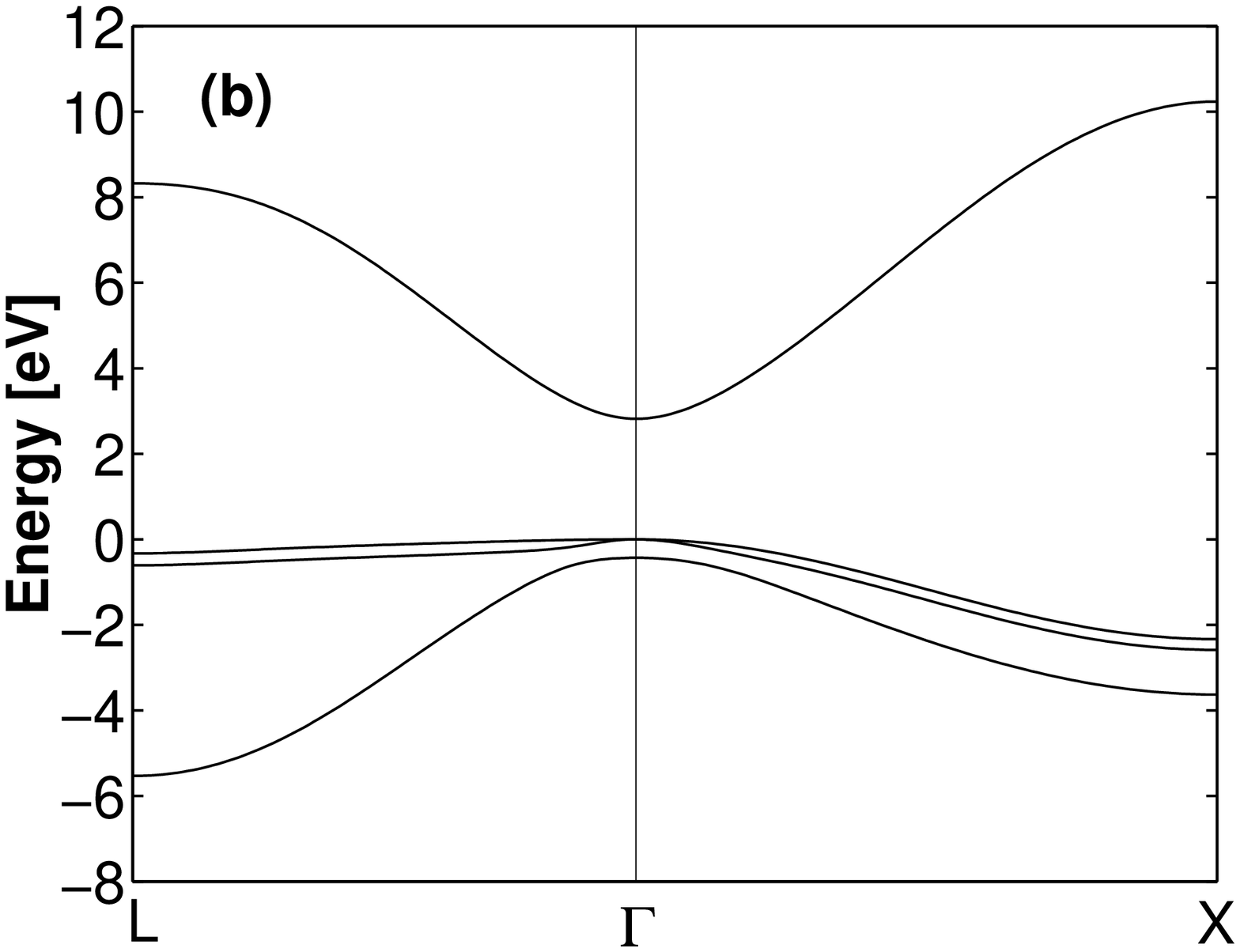}
}
\caption{Tight-binding band structures for CdSe (a) and ZnSe (b)}
\label{fig:bandstructure}       
\end{center}
\end{figure}
The coupling of the basis orbitals is limited to nearest and next
nearest neighbors. Following Ref.~\onlinecite{chadi}, the spin-orbit
component of the bulk Hamiltonian $H^{\text{bulk}}$ couples only $p$
orbitals at the same atom. With the two center approximation of
Slater and Koster~\cite{slaterkoster} we are left with only 8
independent matrix elements $E_{\alpha,\alpha'}(\vec{R}'-\vec{R})_{^{\nu,\nu'}}$.\\
In $\vec{k}$ space, with the basis states
$|\vec{k},\nu,\alpha,\sigma\rangle$, the electronic properties of
the pure bulk material are modelled by an $8\times 8$ matrix
$\mathbf{H}^{\text{bulk}}(\vec{k})$ (for each $\vec{k}$ from the
first Brillouin zone). This matrix depends on the different
TB-parameters $E_{\alpha,\alpha'}(\vec{R}'-\vec{R})_{^{\nu,\nu'}}$.
By analytical diagonalization for special $\vec{k}$ directions, the
electronic dispersion $E_n(\vec{k})$ is obtained as a function of
the TB-parameters; here $n$ is the band index. Equations for the
different TB-parameters
$E_{\alpha,\alpha'}(\vec{R}'-\vec{R})_{^{\nu,\nu'}}$ can now be
deduced as a function of the Kohn-Luttinger-parameters
($\gamma_1$,$\gamma_2$,$\gamma_3$), the energy gap $E_\text{gap}$,
the effective electron mass $m_e$ and the spin-orbit-splitting
$\Delta_{\text{so}}$. The zero level of the energy scale is fixed to
the valence-band maximum. The used material parameters for CdSe and
ZnSe are given in Table~\ref{tab:materialpara}. The resulting
numerical values for the different TB-parameters (obtained by
optimizing them so that the resulting TB band-structure reproduces
the parameters given in Table~\ref{tab:materialpara}) are summarized
in the Table~\ref{tab:TB-parameters} (with and without taking into
account a site-diagonal parameter for the spin-orbit coupling).
Within this approach, the characteristic properties of the band
structure in the region of the $\Gamma$ point are well reproduced,
as can be seen from Fig.~\ref{fig:bandstructure}, which shows the
TB-bands of bulk CdSe and ZnSe (using the TB-parameters with
spin-orbit coupling). When comparing with band structure results
from the literature\cite{cohen}, one sees that the three valence
($p$-) bands are excellently reproduced whereas the $s$-like
conduction band is well reproduced only close to the $\Gamma$-point.
This is understandable, because higher (unoccupied) conduction bands
are neglected, and can be improved by taking into account more basis
states per unit cell. But for a reproduction of the electronic
properties in the region near the $\Gamma$-point, which is important
for a proper description of the optical properties of the
semiconductor materials, the $s_cp_a^3$-TB-model is certainly
sufficient and satisfactory.
\begin{table}
\begin{center}
\begin{tabular}{|c|c|c|c|}
\hline Material & Parameter & TB & TB-NO SO\\\hline\hline
ZnSe & $E_{xx}(000)_{aa}$ & -1.7277 & -2.0413\\
 & $E_{ss}(000)_{cc}$ & 7.0462 & 12.1223 \\
 & $E_{sx}\left(\frac{1}{2}\frac{1}{2}\frac{1}{2}\right)_{ac}$ & 1.1581 &
0.2990 \\
 & $E_{xx}\left(110\right)_{aa}$ & 0.1044 & 0.2185 \\
 & $E_{xx}\left(011\right)_{aa}$ & 0.1874 & 0.0732 \\
 & $E_{xy}\left(110\right)_{aa}$ & 0.3143 & 0.4285 \\
 & $E_{ss}\left(110\right)_{cc}$ & -0.3522 & -0.7752\\
 & $\lambda$ & 0.1433 & 0 \\\hline\hline
CdSe & $E_{xx}(000)_{aa}$ & -1.2738 & -1.7805\\
 & $E_{ss}(000)_{cc}$ &  3.6697 & 10.8053 \\
 & $E_{sx}\left(\frac{1}{2}\frac{1}{2}\frac{1}{2}\right)_{ac}$ & 1.1396 &
0.4260\\
 & $E_{xx}\left(110\right)_{aa}$ & 0.0552 & 0.2161\\
 & $E_{xx}\left(011\right)_{aa}$ & 0.1738 & 0.0129 \\
 & $E_{xy}\left(110\right)_{aa}$ & 0.1512 & 0.3120 \\
 & $E_{ss}\left(110\right)_{cc}$ & -0.1608 & -0.7554 \\
 & $\lambda$ & 0.1367 & 0 \\\hline
\end{tabular}
\caption{TB-parameters (in eV) with (TB) and without (TB-NO SO)
spin-orbit coupling for ZnSe and CdSe, using the notation of
Ref.~\onlinecite{slaterkoster}.}
\label{tab:TB-parameters}
\end{center}
\end{table}

\subsection{TB-Model for Embedded Quantum Dots and Nanocrystals}
Having determined suitable TB-parameters for the bulk materials
(here CdSe and ZnSe) a EQD or NC can be modelled simply by using the
TB-parameters of the bulk materials for those sites (or unit cells)
occupied by atoms (or molecules) of this material. Concerning the
on-site matrix elements this condition is unambiguous. Concerning
the intersite matrix elements one also uses the bulk matrix
elements, if the two sites are occupied by the same kind of
material, but one has to use suitable averages of the bulk intersite
matrix elements for matrix elements over interfaces between
different material, i.e. if the two sites (or unit cells) are
occupied by different atoms (or molecules). Concerning the surfaces
or boundaries of the nanostructure there are different
possibilities. One can use fixed boundary conditions, i.e.
effectively use zero for the hopping matrix elements from a surface
atom to its fictitious nearest neighbors, or (for the embedded QDs)
one can use periodic boundary conditions to avoid any surface
effects, which artificially arise from the finite cell size used for
the EQD-modelling. For the NCs the best thing to do is a realistic,
atomistic modelling of the organic ligands covering the NC-surface,
as described in Refs.\cite{pokrant99,rabani2001,puzder2004-2}.
Within the restricted basis set thus selected the ansatz for an
electronic eigenstate of the EQD or NC is, of course, a linear
combination of the atomic orbitals
$|\nu,\alpha,\sigma,\vec{R}\rangle$:
\begin{equation}
|\Phi\rangle=\sum_{\alpha,\nu,\sigma,\vec{R}}u_{\nu,\alpha,\sigma,\vec{R}}|\nu,\alpha,\sigma,\vec{R}\rangle \; .
\label{eq:zustand}
\end{equation}
Here $\vec{R}$ denotes the unit cell, $\alpha$ the orbital type, $\sigma$ the spin and $\nu$ an anion or cation.
Then the Schr{\"o}dinger equation leads to the following finite matrix
eigenvalue problem:
\begin{equation}
\sum_{\alpha,\nu,\sigma,\vec{R}} \langle{\nu',\alpha',\sigma',\vec{R}'}|H|{\nu,\alpha,\sigma,\vec{R}}\rangle
u_{\nu,\alpha,\sigma,\vec{R}}-Eu_{\nu',\alpha',\sigma',\vec{R}'}=0 \; ,
\label{eq:eigenvalueprob}
\end{equation}
where $E$ is the energy eigenvalue. The shortcut notation
$\langle{\nu',\alpha',\sigma',\vec{R}'}|H|{\nu,\alpha,\sigma,\vec{R}}\rangle=H_{l\vec{R}',m\vec{R}}$ is used
in the following for the
matrix elements with $l=\nu',\alpha',\sigma'$ and $m=\nu,\alpha,\sigma$.

The matrix elements for CdSe and ZnSe without strain are denoted by
$H^{0}_{l\vec{R}',m\vec{R}}$. For these matrix elements the
TB-parameters $E_{\alpha,\alpha'}(\vec{R}'-\vec{R})_{^{\nu,\nu'}}$
of the bulk materials, determined in Sec.~\ref{sec:2a}, are used.
For the off-diagonal matrix elements over interfaces and the
diagonal matrix elements of the selen atoms at the interface between
dot and barrier, which can not unambiguously be referred to belong
to the ZnSe or CdSe, respectively, we choose the mean value of the
parameters for the two materials. Furthermore, a parameter for the
valence-band offset $\Delta E_{V}$ has to be included in the model.
This means that for CdSe in a heterostructure, i.e. surrounded by a
barrier ZnSe material,  all diagonal matrix elements are shifted
just by $\Delta E_{V}$ compared to the bulk CdSe diagonal matrix
elements. In the literature different values for $\Delta E_{V}$ can
be found, they vary in the range of 10\,\%-30\,\% of the band gap
difference between CdSe and
ZnSe~\cite{kurtz2001,yiyang2001,lankes96}. We have performed
calculations with valence-band offsets of $\Delta E_V=0.108$ eV,
$\Delta E_V=0.22$ eV and $\Delta E_V=0.324$ eV, which  corresponds
to 10\,\%, 20\,\% and 30\,\% of the difference of the band gaps. We
find that these different choices for  $\Delta E_{V}$ shift the EQD
energy gap $E^{\text{QD}}_\text{gap}$ by less than 2\,\%. This
shows, that the results are not much affected by the specific choice
of the valence-band offset $\Delta E_V$. Therefore, in the
following, an intermediate  value of $\Delta E_V=0.22$ eV is chosen.

Furthermore, in a heterostructure of two materials with different
lattice constants, strain effects have to be included for a
realistic description of the electronic states, because the distance
between two CdSe unit cells and the bond angles are not the same as
the corresponding equilibrium values in bulk CdSe. This means that
the intersite TB matrix elements $H_{l\vec{R}',m\vec{R}}$ in the EQD
differ from the $H^0_{l\vec{R}',m\vec{R}}$ matrix elements in the
bulk material. In general, a relation
\begin{equation}
H_{l\vec{R}',m\vec{R}}=H^0_{l\vec{R}',m\vec{R}}f(\vec{d}^{\,0}_{\vec{R}'-\vec{R}},
\vec{d}_{\vec{R}'-\vec{R}}) \label{eq:harrisonrule}
\end{equation}
has to be expected, where $\vec{d}^{\,0}_{\vec{R}'-\vec{R}}$ and
$\vec d_{\vec{R}'-\vec{R}}$ are the bond vectors between the atomic
positions of the unstrained and strained material, respectively. The
function $f(\vec{d}^{\,0}, \vec d)$ describes, in general, the
influence of the bond length and the bond angle on the intersite
(hopping) matrix elements. For lack of a microscopic theory for the
functional form we use as a simplified model assumption
$f(\vec{d}^0_{\vec{R}'-\vec{R}},
\vec{d}_{\vec{R}'-\vec{R}})=\left(d^0_{\vec{R}'-\vec{R}}/d_{\vec{R}'-\vec{R}}\right)^2$.
With this $d^{-2}$ ansatz, the interatomic matrix elements
$H_{l\vec{R}',m\vec{R}}$, with $\vec{R}'\neq\vec{R}$, are given by
\begin{equation}
H_{l\vec{R}',m\vec{R}}=H^0_{l\vec{R}',m\vec{R}}\left(\frac{d^0_{\vec{R}'-\vec{R}}}{d_{\vec{R}'-\vec{R}}}\right)^2
\; . \label{eq:harrisonrule1}
\end{equation}
This corresponds to Harrison's~\cite{harrison79} $d^{-2}$ rule, the
validity of which has been demonstrated for II-VI-materials by Sapra
\textit{et al.}~\cite{sapra2002}. More sophisticated ways to treat
the scaling of the interatomic matrix elements, e.g. by calculating
the dependence of energy bands on volume effects and different
exponents for different orbitals, can be found in the
literature~\cite{jancu98,santo,lee2004}. Furthermore the results of
Bertho \textit{et al.}~\cite{bertho91} for the calculations of
hydrostatic and uniaxial deformation potentials in case of ZnSe show
that the $d^{-2}$ rule should be a reasonable approximation. Our
model assumption for the function $f(\vec{d}^0,\vec d)$ means that
we neglect the influence of bond angle distortion. Though energy
shifts due to bond angle distortions have been found for InAs
EQDs\cite{santo}, here the negligence of bond angle distortion can
be justified when exclusively taking into account the coupling
between s- and p-orbitals at nearest neighbor sites. Piezoelectric
fields, which are usually considered to be less important for the
zinc blende structures realized in CdSe and
ZnSe~\cite{fonoberov2003}, are also not taken into account in our
model.

The problem is now reduced to the diagonalization of a finite but very large
matrix. To calculate the eigenvalues of this matrix, in particular the bound electronic states in the
QD, the folded spectrum method~\cite{zunger94} is applied to the
eigenvalue problem of Eq.~\ref{eq:eigenvalueprob}.

\section{Results for a pyramidal ${\bf CdSe}$ Embedded Quantum Dot}
\label{sec:resultspyramida}
\subsection{Geometry and Strain}
\begin{figure}[t]
\begin{center}\resizebox{0.5\textwidth}{!}{%
  \includegraphics{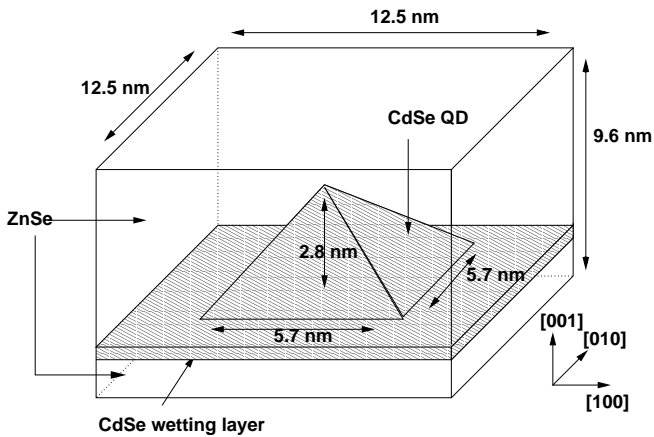}
}
\caption{Schematic visualization of the pyramidal CdSe QD buried in a ZnSe matrix. The wetting layer has a thickness of
one lattice constant ($1\,a$) of bulk ZnSe. The pyramidal QD has a base length $b$ of ten times the
ZnSe lattice constant ($b=10\,a$).}
\label{fig:geometry}       
\end{center}
\end{figure}

To model a CdSe QD embedded into a ZnSe barrier material we choose a
finite (zinc blende) lattice within a box with fixed boundary
conditions. Within this box we consider a CdSe WL of thickness $1 a$
(lattice constant of the conventional unit cell, i.e. about two
anion and two cation layers), and on top of this wetting layer there
is a pyramidal QD with base length $b$ and height $h=b/2$. For the
matrix elements corresponding to sites within the WL or the QD we
choose the TB-values appropriate for CdSe, for all other sites
within the box the ones for ZnSe. Figure~\ref{fig:geometry} shows a
schematic picture of this geometry we use to model the EQD. We
investigate EQDs with a base lengths $b$ of $6\,a$, $8\,a$ and
$10\,a$, where $a=5.668\,\textrm{\AA}$ is the lattice constant of
the bulk ZnSe material. Cells with the dimensions of
$18\,a\times18\,a\times15\,a$ (38 880 atoms),
$20\,a\times20\,a\times16\,a$ (51 200 atoms) and
$22\,a\times22\,a\times17\,a$ (65 824 atoms) are used for the
calculations.
Figure~\ref{fig:geometry} shows the EQD with a base length $b$ of
$10\,a$. Fixed boundary conditions are applied to avoid a dot-dot
coupling in contrast to periodic boundary conditions~\cite{santo}.
The total size of the cells is chosen so that the boundary
conditions affect the energy gap of the EQD by less than 2\,\%.

To consider strain effects in our model the knowledge of the strain
tensor $\boldsymbol{\epsilon}$ is necessary. The strain tensor
$\boldsymbol{\epsilon}$ is related to the strain dependend
relative atomic positions $\vec{d}_{\vec{R}'-\vec{R}}$ by
\begin{equation}
\vec{d}_{\vec{R}'-\vec{R}}=(\mathbbm{1}+\boldsymbol{\epsilon})\vec{d}^{\,\,0}_{\vec{R}'-\vec{R}}\;
.
\end{equation}
\begin{figure}[b]
\begin{center}
\resizebox{0.4\textwidth}{!}{%

  \includegraphics{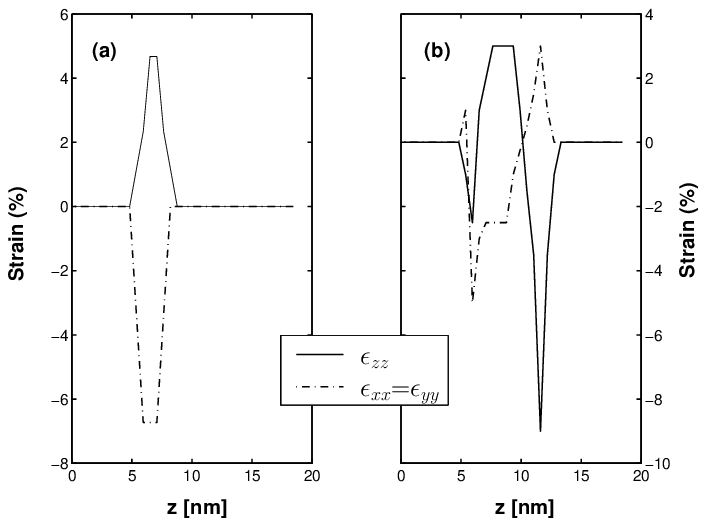}
} \caption{Strain distribution in and around the embedded pyramidal
CdSe QD with a base length of $b=10\,a$. The WL at the base of the
QD is $1\,a$ thick. The whole structure is buried in a ZnSe matrix.
Line scans along the [001] direction through the WL outside the dot
(a) and inside the dot through the tip of the pyramid (b) are
displayed. The diagonal elements of the strain tensor
$\boldsymbol{\epsilon}$ are shown as solid ($\epsilon_{zz}$) and
dashed-dotted lines ($\epsilon_{xx}=\epsilon_{yy}$).}
\label{fig:strainprof}       
\end{center}
\end{figure}
To appoint the strain tensor outside the EQD, the WL is treated as a
quantum film. In the absence of a shear strain
($\epsilon_{i,j}\sim\delta_{i,j}$) for a coherently grown film, the
strain components are given by~\cite{chuang95}
\begin{eqnarray}
\epsilon_{||}=\epsilon_{xx}=\epsilon_{yy}&=&\frac{a_S-a_D}{a_D}\\
\epsilon_{\bot}=\epsilon_{zz}&=&-\frac{C_{12}}{C_{11}}\epsilon_{||}\; .
\end{eqnarray}
Here $a_D$ is the lattice constant of the unstrained film material
and $a_S$ denotes the parallel lattice constant of the substrate. In
Table~\ref{tab:materialpara} the cubic elastic constants $C_{ij}$ of
the bulk materials are given. The resulting strain profile for a
line scan in z-direction outside the dot is shown in
Fig.~\ref{fig:strainprof} (a). In Ref.~\onlinecite{grundmann98}
Stier \textit{et al.} considered a similar strain profile for an
InAs/GaAs EQD. The lattice mismatch of approximately 7\,\% in the
InAs/GaAs system is nearly the same as for the CdSe/ZnSe system. So
our calculated strain profile shows the same behavior as the profile
in Ref.~\onlinecite{grundmann98} for a line scan in z-direction
outside the EQD.

To obtain the strain profile inside the EQD we use a model strain
profile, which shows a similar behavior as the strain profiles which
are given in Refs.~\onlinecite{pryor98, grundmann98} for a line scan
in z-direction through the tip of the  pyramid. This model strain
profile is displayed in Fig.~\ref{fig:strainprof} (b). The shear
components, $\epsilon_{xy}$, $\epsilon_{xz}$ and $\epsilon_{yz}$,
can be neglected, at least away from the boundaries of the
dot~\cite{grundmann1}.

\subsection{Bound single particle states}

\begin{figure}[t]
\begin{center}
\resizebox{0.5\textwidth}{!}{%
  \includegraphics{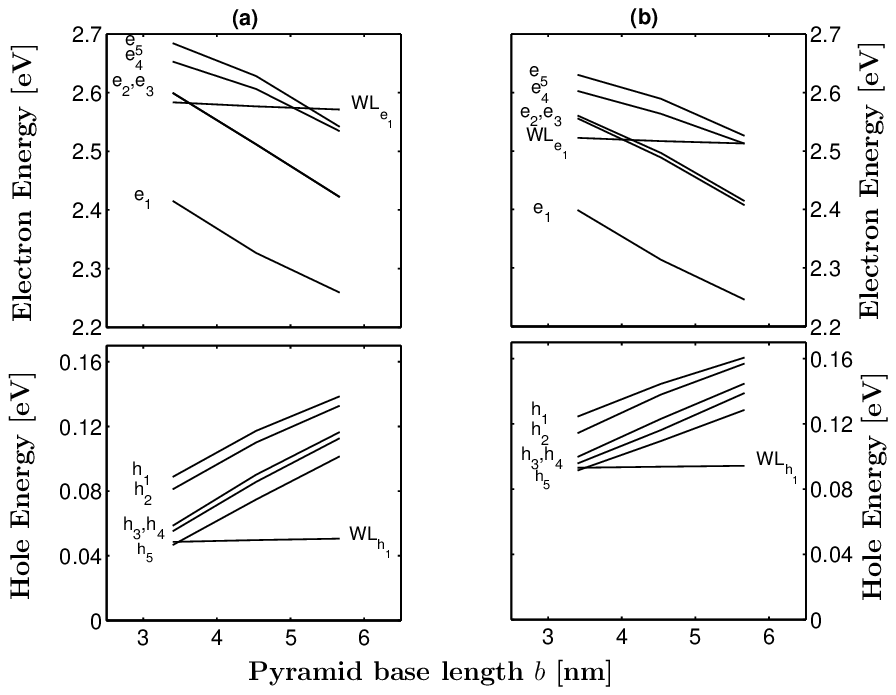}
} \caption{Electron and hole  energies for embedded pyramidal
CdSe/ZnSe QDs on a WL of thickness $1 a$ (roughly 4 monolayers) for
different base length of the dot without strain (a) and with strain
effects taken into account (b). The ground state energies for
 electrons ($\text{WL}_{e_1}$) and holes ($\text{WL}_{h_1}$) for the WL (of
thickness $1 a$) alone are also displayed. }
\label{fig:verglbound}       
\end{center}
\end{figure}

We have calculated the first five states for electrons and holes for
three different EQD sizes. These calculations are done with and
without including strain effects. For the evaluations without strain
we have chosen the exponent in Eq.~(\ref{eq:harrisonrule1}) to be
zero. The energy spectrum obtained from these calculations is shown
in Fig.~\ref{fig:verglbound} (a) without strain and in
Fig.~\ref{fig:verglbound} (b) including strain effects. The states
are labeled by $e_1$ and $h_1$ for electron and hole ground states,
$e_2$ and $h_2$ for the first excited states, and so on. All
energies are measured relative to the valence-band maximum of  ZnSe.
Figure~\ref{fig:verglbound} also shows the size dependence of the
electron and hole energy levels. The energies are compared to the
ground state energies for electrons and holes in the $1\,a$ thick
CdSe WL ($\text{WL}_{e_1}$ and $\text{WL}_{h_1}$, respectively),
which is calculated separately for a coherently strained quantum
film (i.e. the WL without the QD). As expected from a naive particle
in a box picture, the binding of electrons and holes becomes
stronger in the EQD when the dot size is increased. The quantum
confinement causes the number of bound states to decrease when the
dot size is reduced. For the EQDs with a base length $b=8\,a$ and
$b=10\,a$ the calucated hole states are well above the WL energy
($\text{WL}_{h_1}$). This is valid for the strain-unaffected and
strained EQD. For the system with $b=6\,a$ we obtain at least four
bound hole states in both models. The energy splitting between the
different states is only slightly influenced by the strain.
Furthermore we see from Fig.~\ref{fig:verglbound} that the number of
bound electron states is influenced by the strain. For the system
with a base length of $b=10\,a$ we get at least three bound-electron
states when we take strain effects into account
(Fig.~\ref{fig:verglbound} (b)). Without strain effects at least 5
bound states are found. So the confinement potential for the
electrons is effectively
reduced by the strain.\\
The bound electron states $e_2$ and $e_3$ are energetically not
degenerate even without strain. This arises from the $C_{2v}$
symmetry of the system. Already from the geometry of the EQD-system
it is clear that there is no $(001)$ mirror plane. Furthermore, if
one considers a $(001)$-plane with sites occupied by Se anions, the
nearest neighbor (cation) planes in $\pm z$-direction are not
equivalent, as in the zinc blende structure the nearest neighbors
above the plane are found in $[111]$-direction and below the plane
in $[1\bar{1}\bar{1}]$-direction. So also for crystallographic
reasons a $(001)$-plane is not a mirror plane. Finally, if one
considers the base plane of the EQD (or the WL) to be this anion
$(001)$-plane, there are different cations, namely Cd above and Zn
below this plane. Therefore, the QD-system has reduced
$C_{2v}$-symmetry. In theories based on continuum models, e.g.
effective mass approximations~\cite{wojs96,grundmann1}, the
discussed effects cannot be accounted for.
\begin{figure*}
\centering
\begin{tabular}{c@{\qquad}c@{\qquad}c@{\qquad}c@{\qquad}c}
\multicolumn{5}{c}{\textbf{Electron states for the $b=10\,a$ base-length dot}}\\
\hline\hline
\multicolumn{5}{c}{With Strain}\\
\hline
$e_1$ & $e_2$ & $e_3$ & $e_4$ & $e_5$ \\
  \resizebox{0.15\textwidth}{!}{\includegraphics{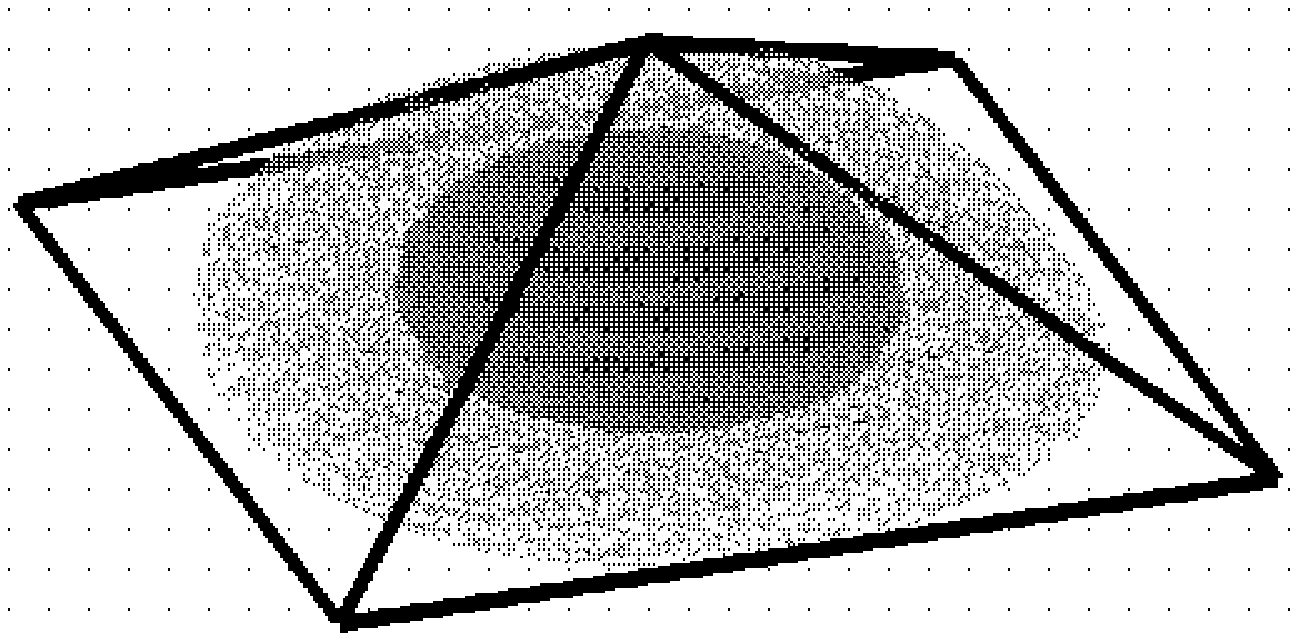}} &
  \resizebox{0.15\textwidth}{!}{\includegraphics{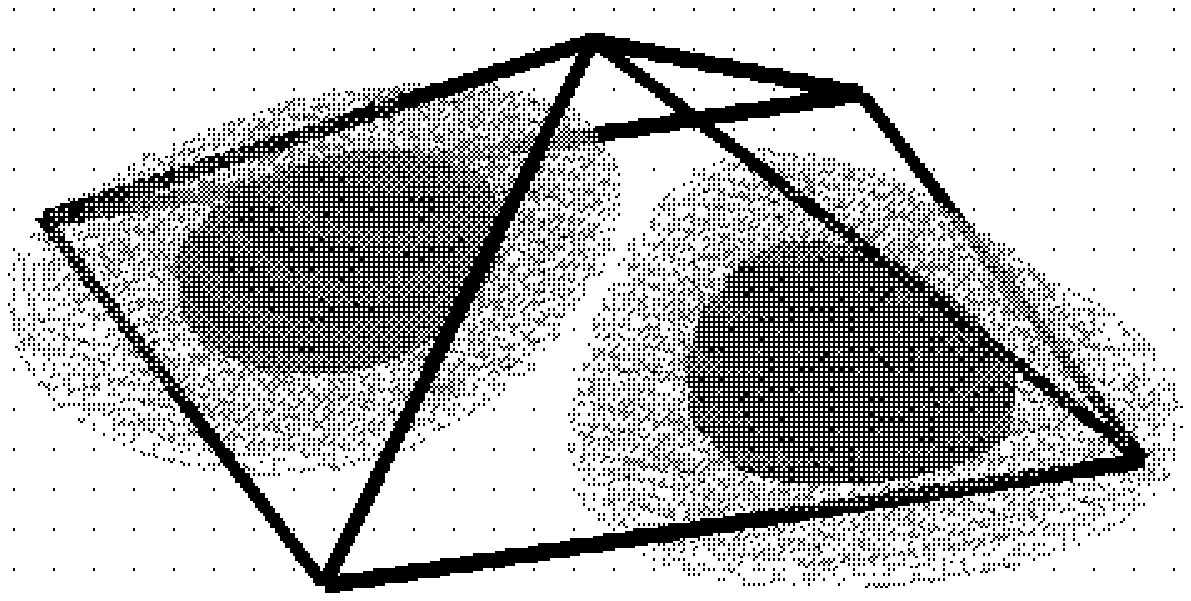}} &
  \resizebox{0.15\textwidth}{!}{\includegraphics{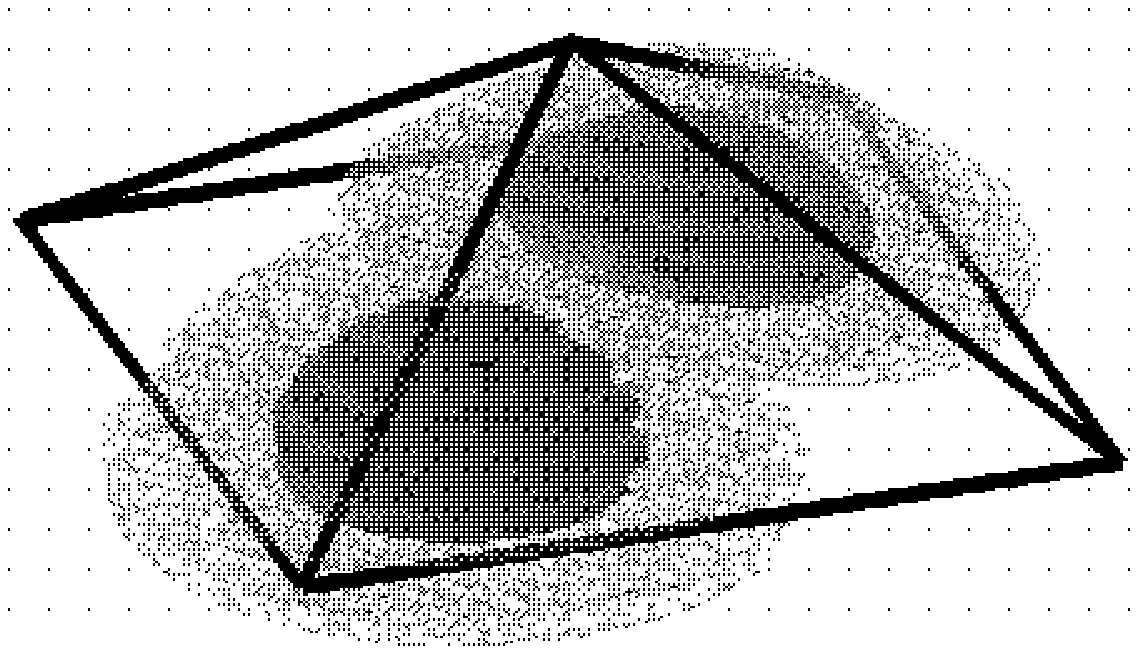}} &
  \resizebox{0.15\textwidth}{!}{\includegraphics{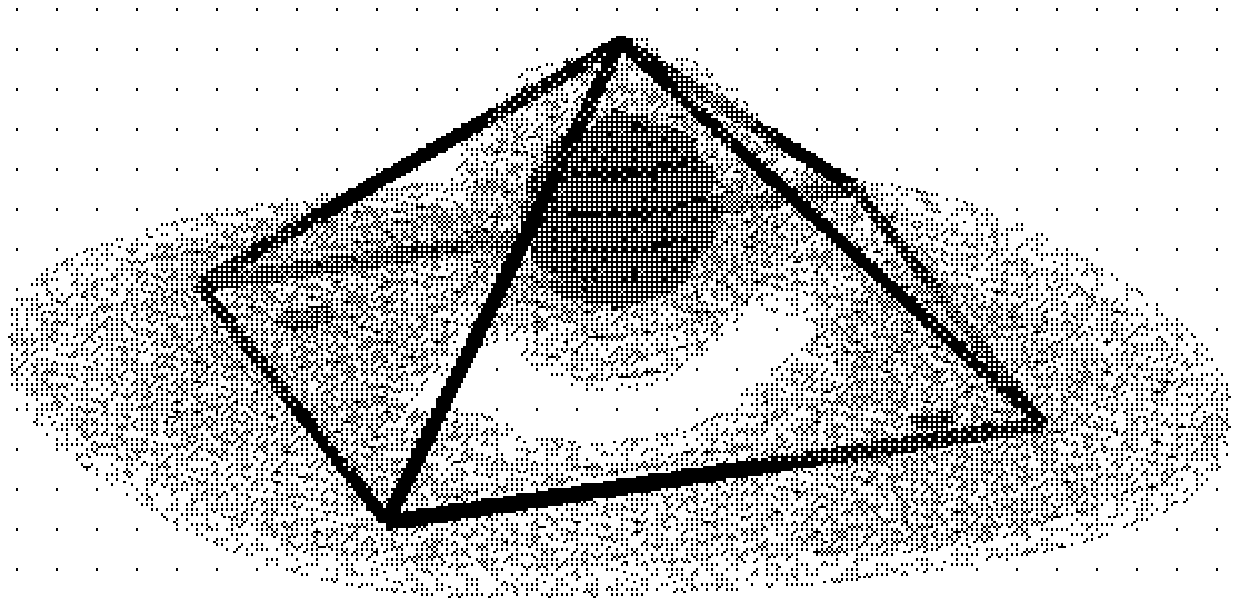}} &
  \resizebox{0.15\textwidth}{!}{\includegraphics{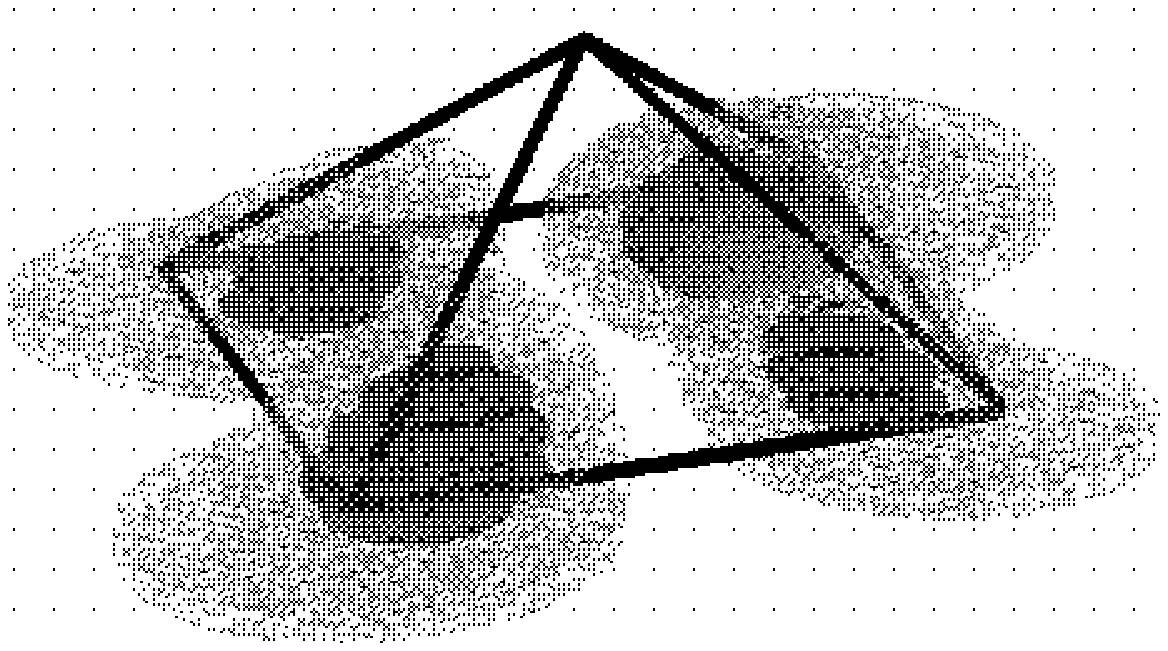}} \\
  \\
\multicolumn{5}{c}{Without Strain}\\
\hline
$e_1$ & $e_2$ & $e_3$ & $e_4$ & $e_5$ \\
  \resizebox{0.15\textwidth}{!}{\includegraphics{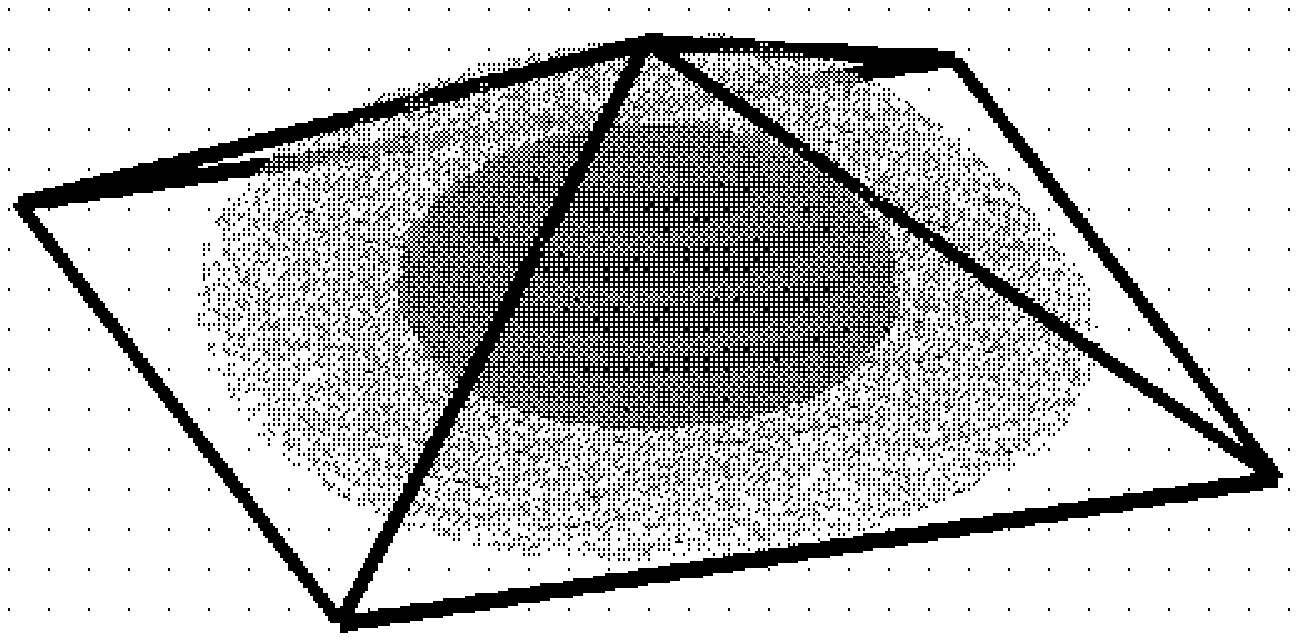}} &
  \resizebox{0.15\textwidth}{!}{\includegraphics{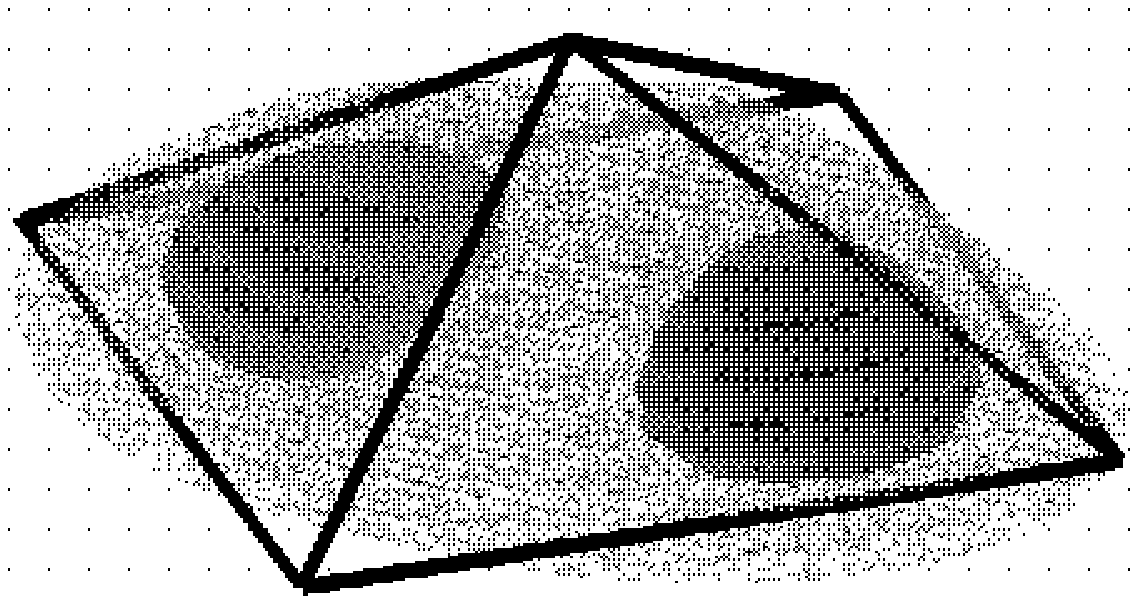}} &
  \resizebox{0.15\textwidth}{!}{\includegraphics{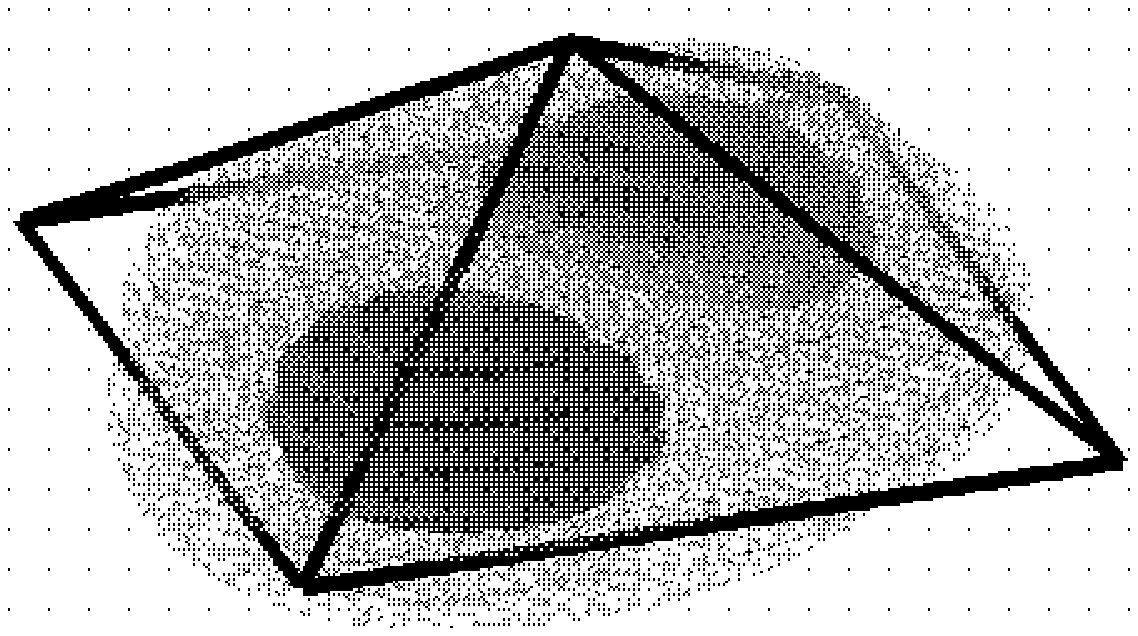}} &
  \resizebox{0.15\textwidth}{!}{\includegraphics{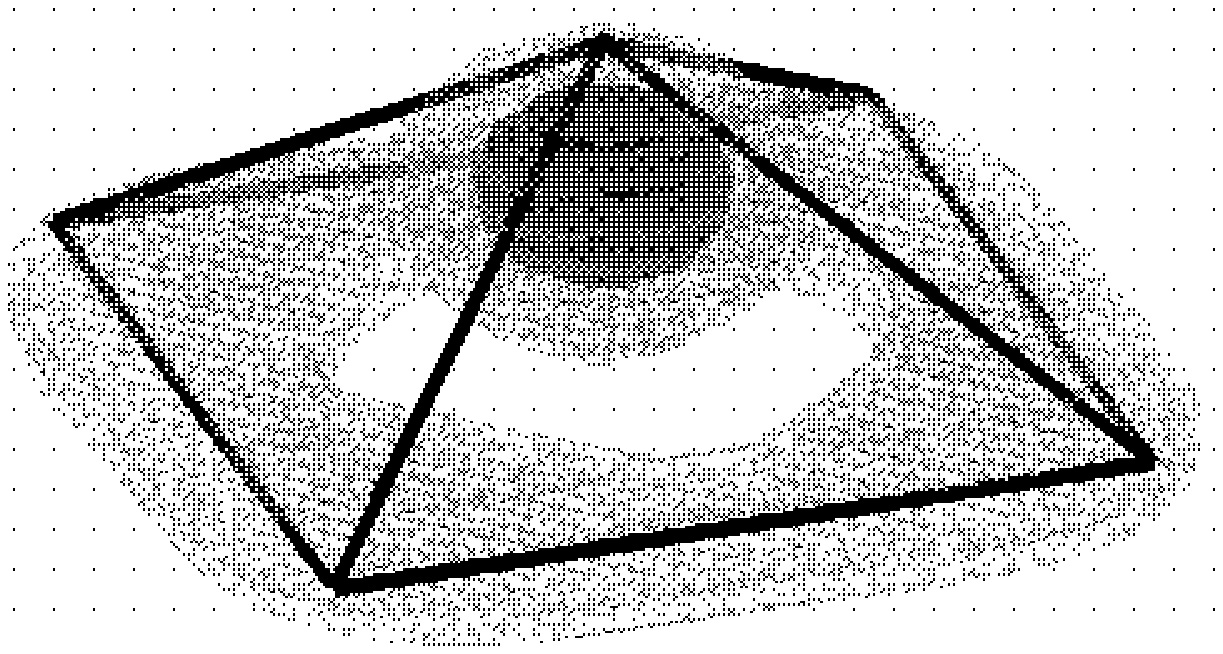}} &
  \resizebox{0.15\textwidth}{!}{\includegraphics{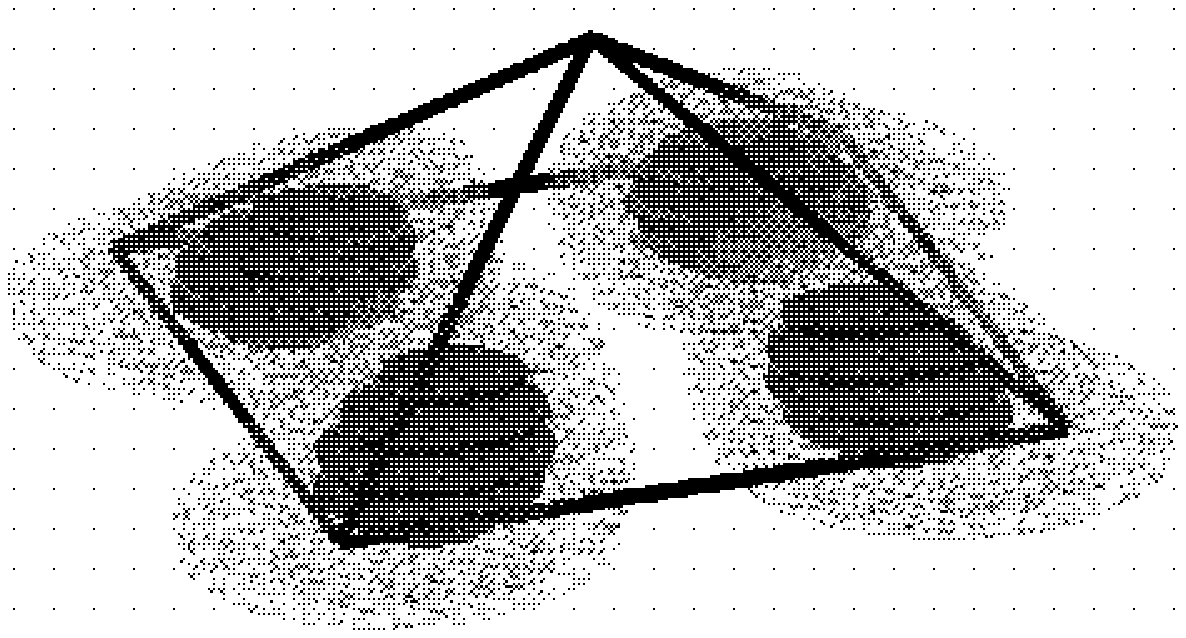}} \\
  \\
  \hline\hline
\end{tabular}
\caption{Isosurfaces of the squared electron wavefunctions with and
without strain for the embedded $b=10\,a$ pyramidal QD. The light
and dark surfaces correspond to 0.1 and 0.5 of the maximum
probability density, respectively.}
\label{fig:electronstates}       
\end{figure*}
\begin{figure*}
\centering
\begin{tabular}{c@{\qquad}c@{\qquad}c@{\qquad}c@{\qquad}c}
\multicolumn{5}{c}{\textbf{Hole states for the $b=10\,a$ base-length dot}}\\
\hline\hline
\multicolumn{5}{c}{With Strain}\\
\hline
$h_1$ & $h_2$ & $h_3$ & $h_4$ & $h_5$ \\
  \resizebox{0.15\textwidth}{!}{\includegraphics{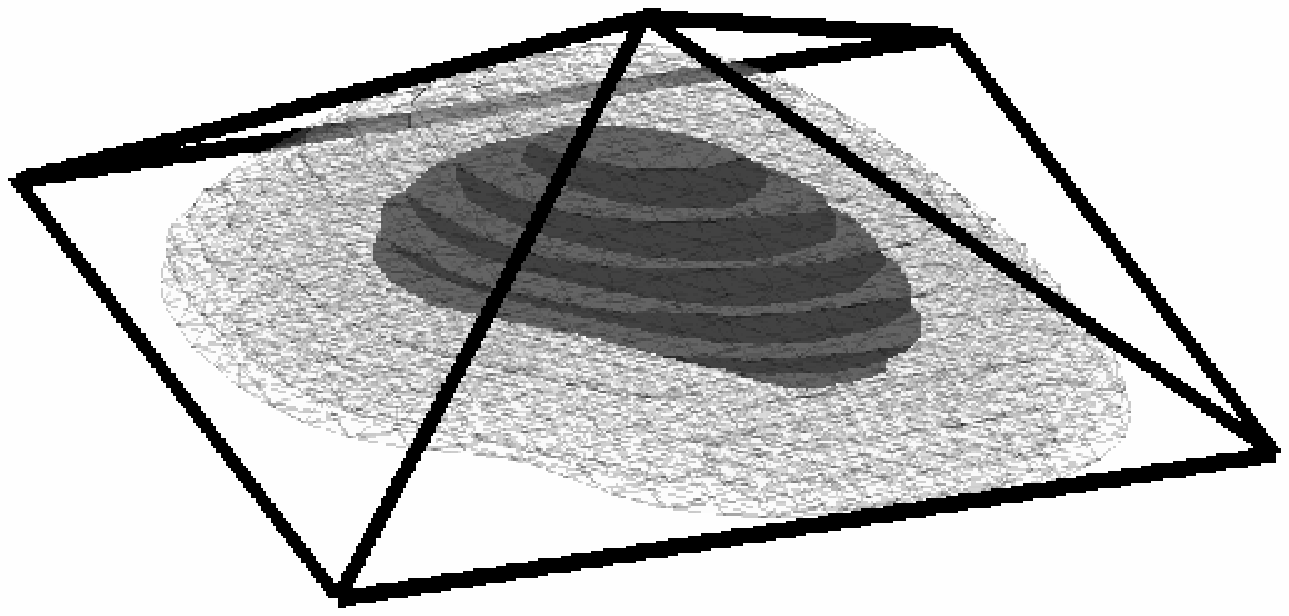}} &
  \resizebox{0.15\textwidth}{!}{\includegraphics{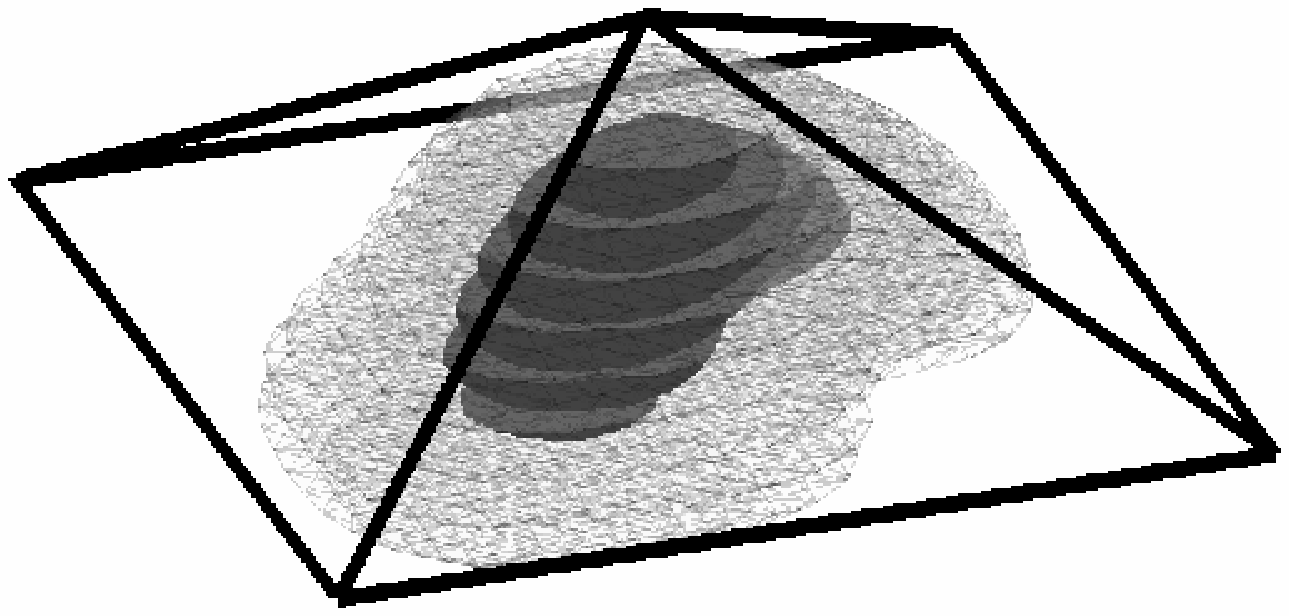}} &
  \resizebox{0.15\textwidth}{!}{\includegraphics{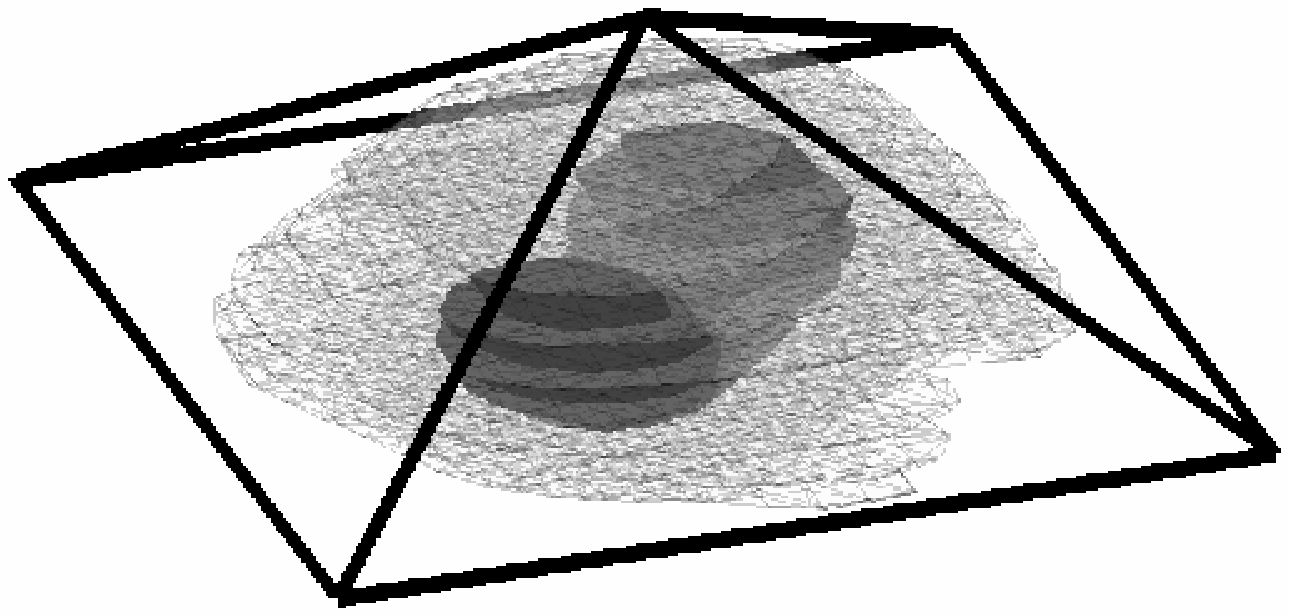}} &
  \resizebox{0.15\textwidth}{!}{\includegraphics{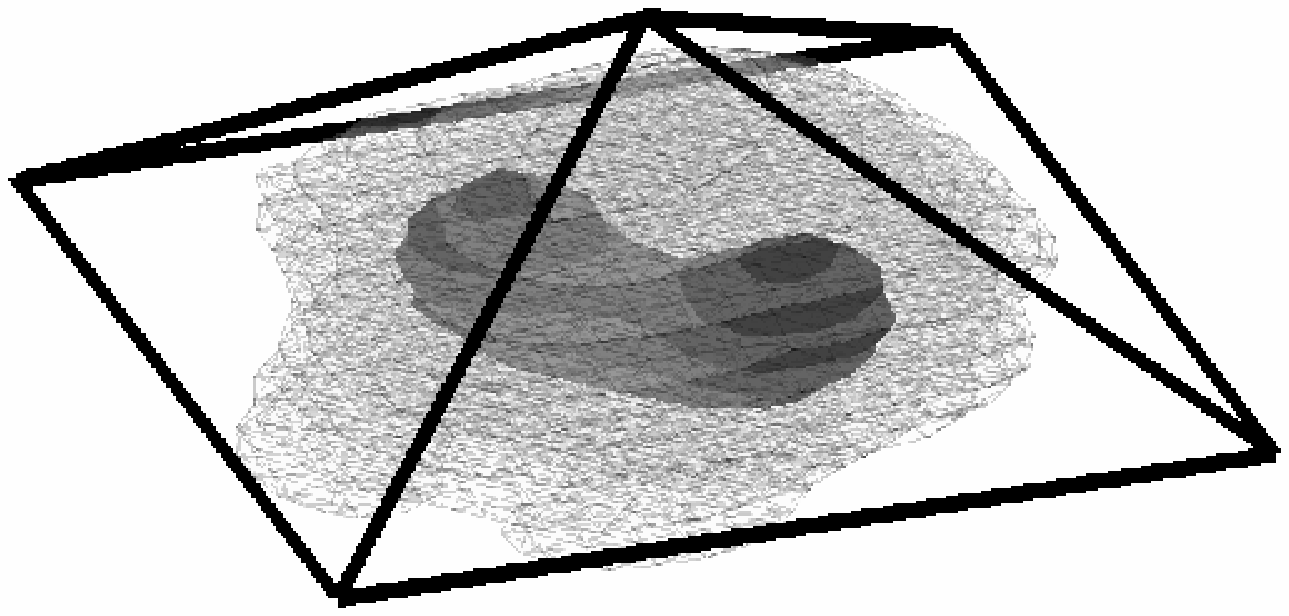}} &
  \resizebox{0.15\textwidth}{!}{\includegraphics{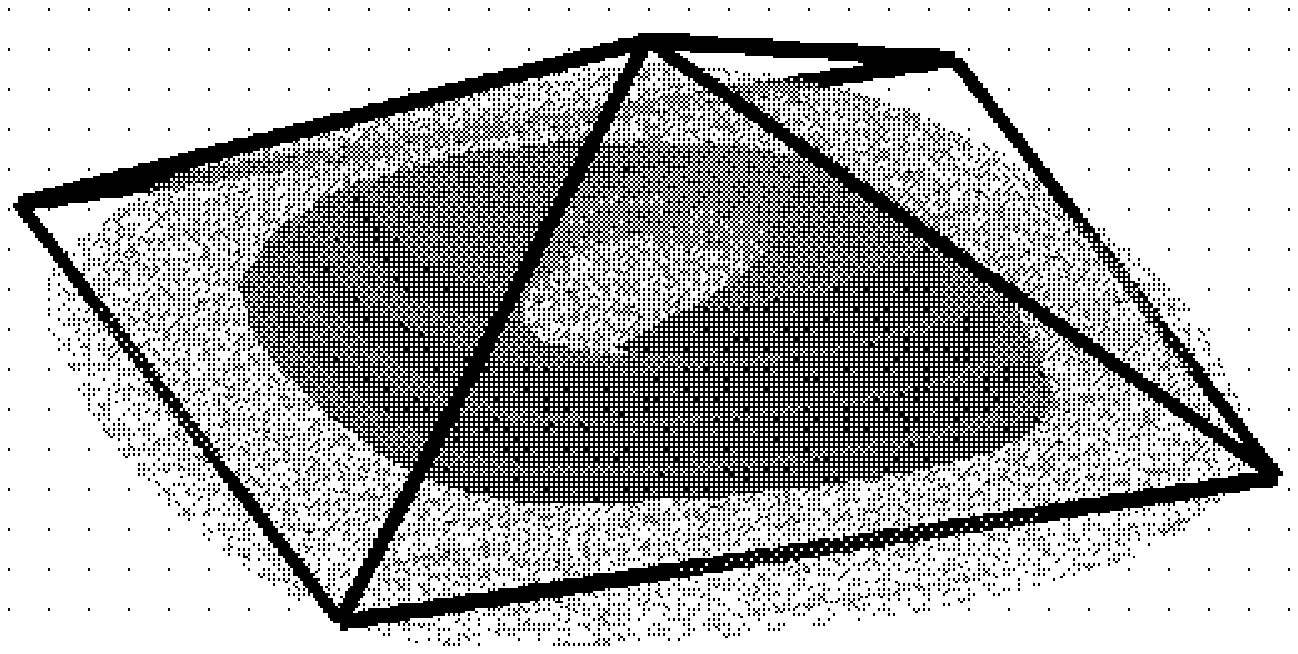}} \\
  \\
\multicolumn{5}{c}{Without Strain}\\
\hline
$h_1$ & $h_2$ & $h_3$ & $h_4$ & $h_5$ \\
  \resizebox{0.15\textwidth}{!}{\includegraphics{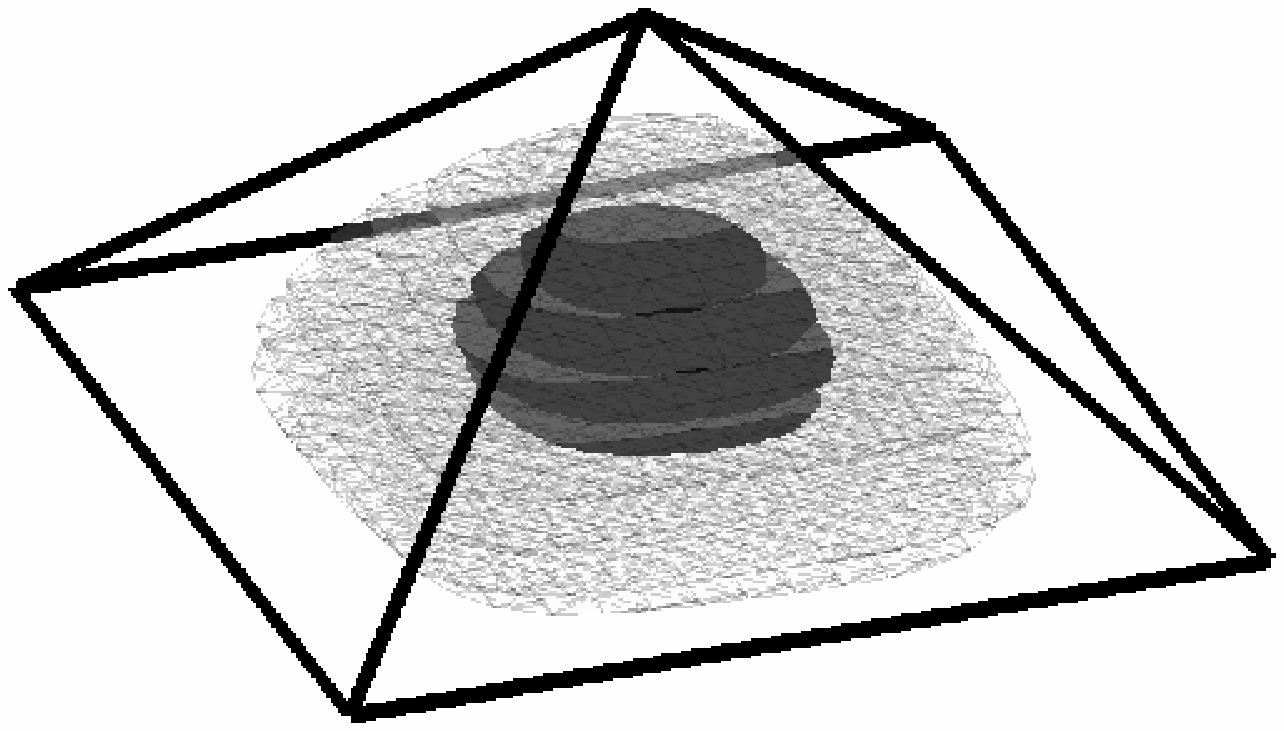}} &
  \resizebox{0.15\textwidth}{!}{\includegraphics{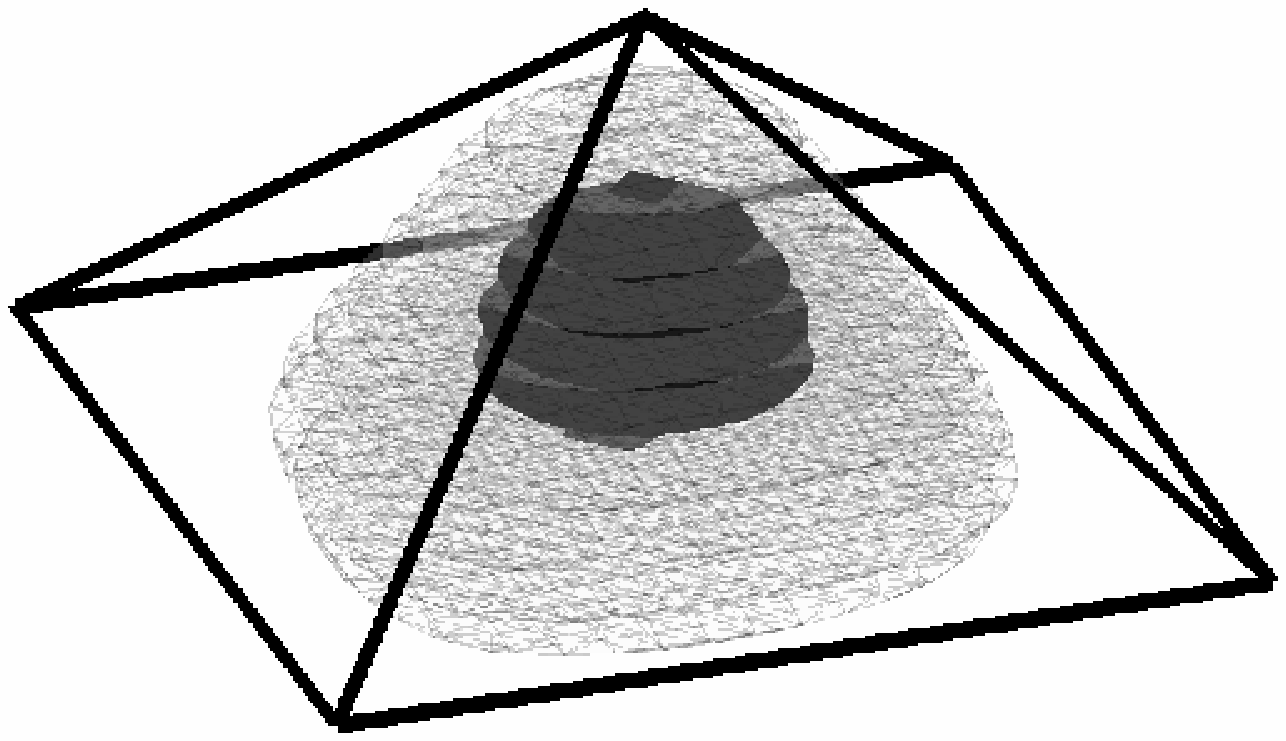}} &
  \resizebox{0.15\textwidth}{!}{\includegraphics{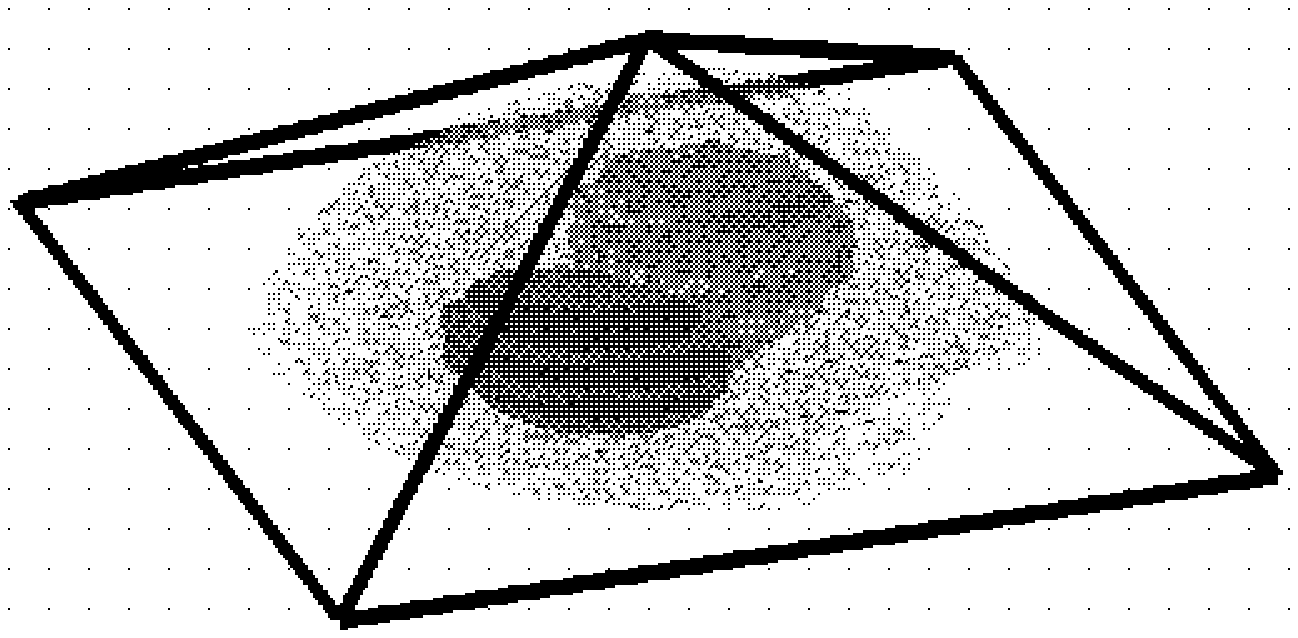}} &
  \resizebox{0.15\textwidth}{!}{\includegraphics{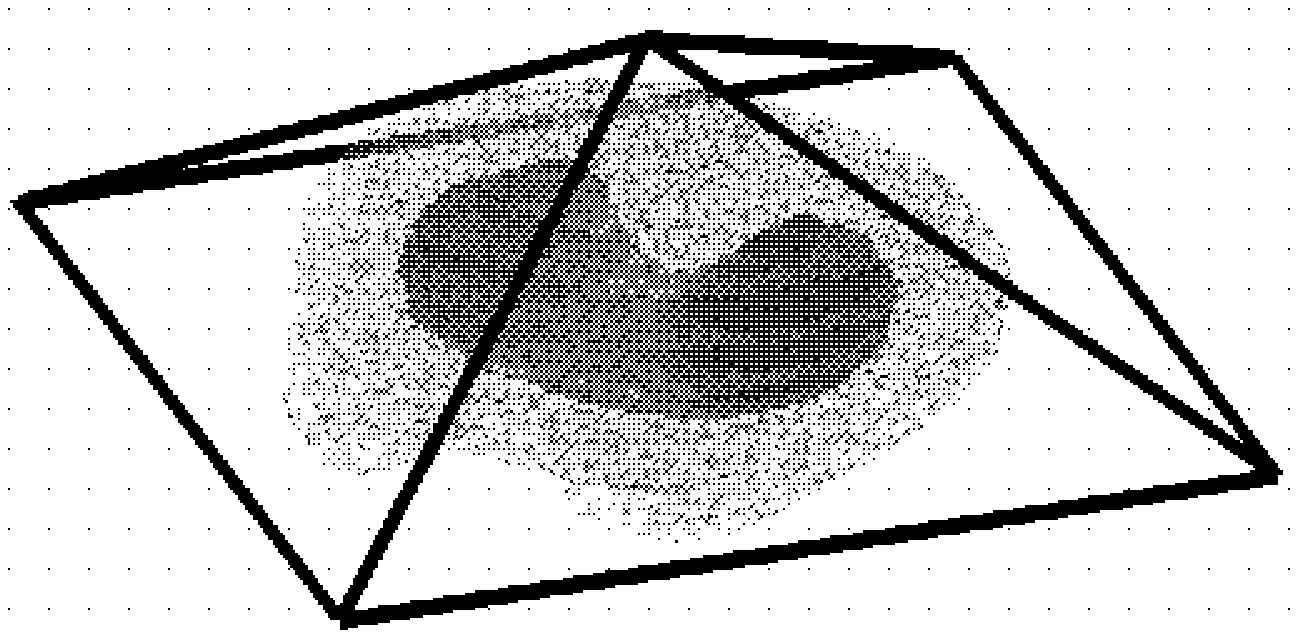}} &
  \resizebox{0.15\textwidth}{!}{\includegraphics{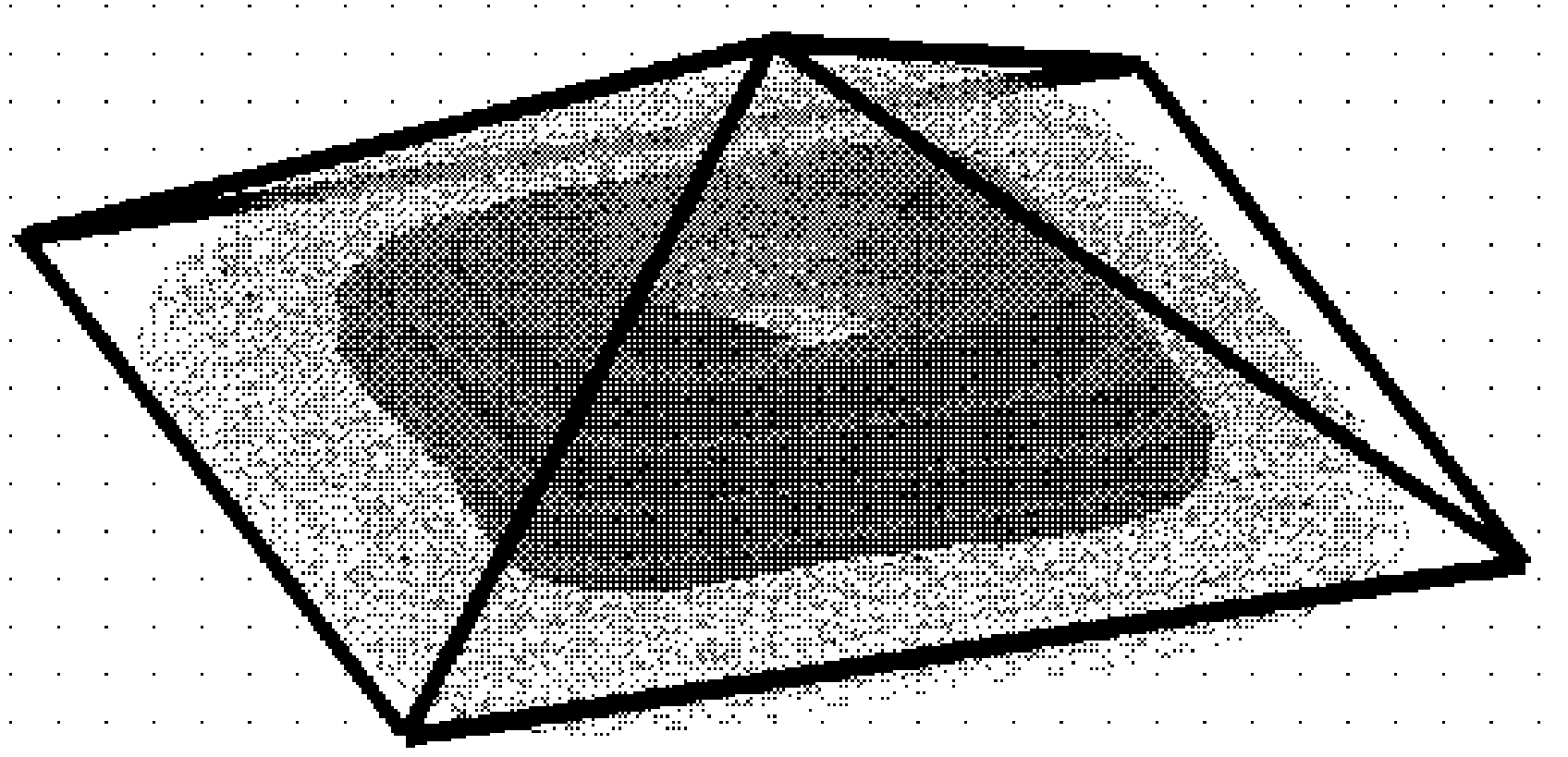}} \\
  \\
  \hline\hline
\end{tabular}
\caption{Isosurfaces plots of the squared hole wave functions with
and without strain for the embedded $b=10\,a$ pyramidal QD. See
caption of Fig.~\ref{fig:electronstates} for more details.}
\label{fig:holestates}       
\end{figure*}
These interfacial effects also affect the one particle wave
functions in the system. In Fig.~\ref{fig:electronstates} the
isosurfaces for the squared electron wavefunctions
$|\Phi_{i}(\vec{r})|^2$ are displayed with and without strain,
respectively. The light and dark isosurface levels are selected as
0.1 and 0.5 of the maximum probability density, respectively. For
both calculations, the lowest electron state $e_1$ is an $s$-like
state according to its nodal structure. The next two states $e_2$
and $e_3$ are $p$-like states. These states are oriented along the
$[1\bar{1}0]$ and the $[110]$ direction, respectively. Due to the
different atomic structure along these directions we find a
$p$-state splitting $\Delta^{0}_{e_2,e_3}=E_{e_3}-E_{e_2}$ for the
unstrained EQD of about $0.43\,$meV. In conventional
$\vec{k}\cdot\vec{p}$ models~\cite{grundmann1,cpryor98} an
unstrained, square-based pyramidal EQD is modelled with a $C_{4v}$
symmetry. In our microscopic model the resulting degeneracy is
lifted and a  splitting occurs as a consequence of the reduction of
$C_{4v}$ symmetry to a $C_{2v}$ zinc blende symmetry.

The strain splits the states $e_2$ and $e_3$ further. Due to the
different atomic structure, the strain profile within each  plane
(perpendicular to the growth z-direction) along the $[110]$ and
$[1\bar{1}0]$ direction is different~\cite{pryor98}. This effect
contributes also to the anisotropy. Due to the fact, that the base
is larger than the top, there is a gradient in the strain tensor
between the top and the bottom of the pyramid. In the EQD, the
cation neighbors above each anion are found in $[111]$ direction
while the cation neighbors below are found in
$[1\bar{1}\bar{1}]$-direction. Therefore, the cations along the
$[1\bar{1}0]$ direction are systematically more stressed than the
cations along the $[110]$ direction. In case of strain we find a
$p$-state splitting of \mbox{$\Delta^{\text{strain}}_{e_2,e_3}=7.1$
meV}. Compared to the states $e_2$ and $e_3$ of the unstrained EQD,
the two lumps of the light isosurfaces are well separated. The
states $e_2$ and $e_3$
reveal nodal planes along the $[110]$ and $[1\bar{1}0]$ direction, respectively.\\
The state $e_4$ for the strained dot is resonant with a WL-state, so
the wave function is leaking into the WL. Also the wave function of
the state $e_5$ is localized at the base of the pyramid but clearly
shows already a finite probability density inside the WL. The states
$e_4$ and $e_5$ of the unstrained EQD are still mainly localized
inside the dot. The classification of the state $e_4$ by its nodal
structure is difficult. $e_4$ is similar to a $p$-state which is
oriented along the $[001]$ direction. The electron state $e_5$ is
$d$-like.\par

Figure~\ref{fig:holestates} shows the isosurface plots of the
squared wavefunction $|\Phi_i(\vec{r})|^2$ for the lowest five hole
states $h_1$-$h_5$ with and without strain. The light and dark
isosurface levels are again selected as 0.1 and 0.5 of the maximum
probability density, respectively. Our atomistic calculation shows
that the hole states cannot be classified by $s$-like ($h_1$),
$p$-like ($h_2$ and $h_3$) or $d$-like ($h_5$) shape according to
their nodal structures. With and without strain the hole states
underly a strong band mixing. So the calculated hole states show no
nodal structures. Therefore the assumption of a single heavy-hole
valence-band for the description of the bound hole states in a EQD
even qualitatively yields incorrect results. In contrast to quantum
well systems, the light-hole and heavy-hole bands are strongly mixed
in a EQD. This result is in good agreement with other multiband
approaches~\cite{grundmann98,fonoberov2003,cpryor98,wang99,wang2000,williamson2000,kim98}.

From Fig.~\ref{fig:holestates} we can also estimate the influence of strain on the different hole states. Without
strain the states $h_1$ and $h_2$ are only slightly elongated along the $[1\bar{1}0]$ and
$[110]$ direction, respectively. Due to strain these states are clearly elongated along these directions. The
states $h_3$-$h_5$ are only slightly affected by strain.

\begin{figure}[t]
\begin{center}
\resizebox{0.5\textwidth}{!}{%
  \includegraphics{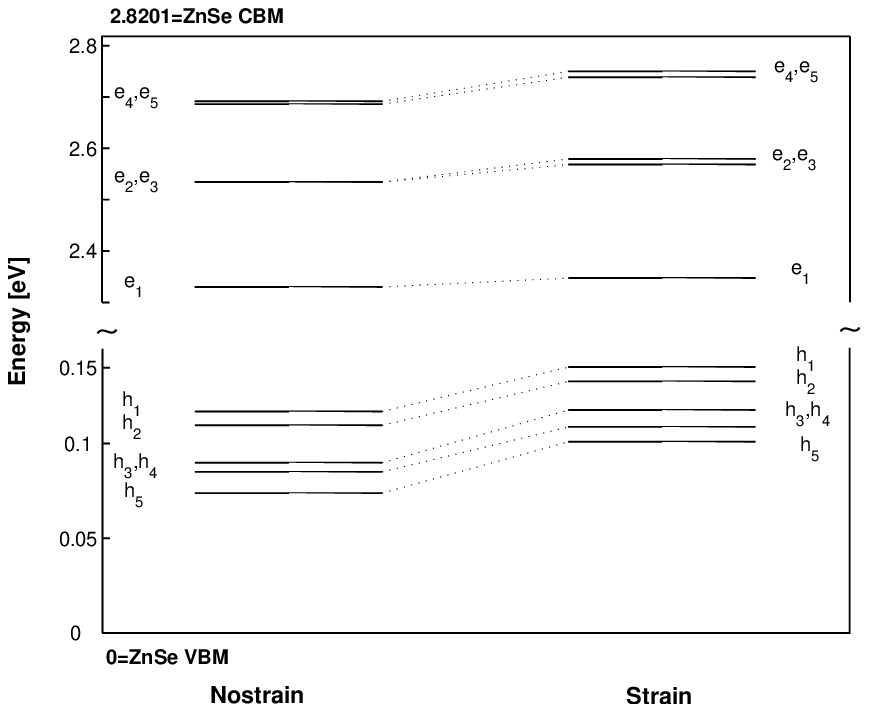}
} \caption{First five electron and hole state energies for the
embedded pyramidal CdSe QD with $b=10\,a$ and a one monolayer (1 ML)
thick WL. On the left-hand side the results for the unstrained EQD
are shown while on the right-hand side the results for the strained
EQD are displayed. The zero of the energy scale is the bulk ZnSe
valence-band maximum (VBM). The energies are compared with the
conduction-band minimum (CBM) of the bulk ZnSe.}
\label{fig:vergl1ML}       
\end{center}
\end{figure}

Another interesting result  is that strain effects shift the
electron states to lower energies and the hole states to higher
energies as displayed in Fig.~\ref{fig:verglbound}.
Figure~\ref{fig:verglbound} also reveals that the WL ground-state
for electrons and holes is shifted in a similar way due to strain.
We observe that strain decreases the EQD gap
$E^{\text{QD}}_\text{gap}=E_{e_1}-E_{h_1}$ by about 1.4\%, lowering
it from the strain-unaffected value 2.12 eV to the value  2.09 eV.
For a biaxial compressive strain in a zinc blende structure, the
conduction-band minimum of a bulk material is shifted to higher
energies while the energy shift of the valence-band maximum depends
on the magnitude of the hydrostatic and shear deformation
energies~\cite{chuang95}. So one would expect that the electron
states are shifted to higher energies due to the fact that CdSe is
compressively strained in the ZnSe-Matrix. This is in contradiction
to the behavior we observe here. To investigate the influence of the
WL states on the one-particle spectrum we use the same model
geometry as shown in Fig.~\ref{fig:geometry} but with a considerably
smaller WL thickness of only one monolayer (ML). A 1 ML thick WL was
also used  before by Santoprete \textit{et al.}~\cite{santo}, Stier
\textit{et al.}~\cite{grundmann98} and Wang \textit{et
al.}~\cite{wang99} for an InAs/GaAs EQD. Figure~\ref{fig:vergl1ML}
shows the comparison of the results for a strain-unaffected and a
strained pyramidal CdSe EQD with a 1 ML thick WL and a base length
of $b=10\,a$.

On the left-hand side of Fig.~\ref{fig:vergl1ML} the first five
electron and hole-state energies for an unstrained EQD are displayed
while the right-hand side shows the energies for the strained EQD.
For a 1 ML thick WL the lowest electron state is, by strain effects,
shifted to higher energies. This is what one would expect for
biaxial compression of the bulk material. Furthermore the splitting
of the $p$-like states $e_2$ and $e_3$ is larger compared to the
results for a $1\,a$ thick WL. The splitting $\Delta^0_{e_2,e_3}$ of
the unstrained EQD with a $1\,a$ thick WL is
$\Delta^0_{e_2,e_3}=0.43$ meV whereas for the system with a 1 ML
thick WL one has \mbox{$\Delta^0_{e_2,e_3}=0.5$ meV}. So the
spltting $\Delta^0_{e_2,e_3}$ is increased by about 16\,\%. With
strain-effects, the splitting for the system with 1 ML WL thickness
$\Delta^{\text{strain}}_{e_2,e_3}= 10.9$ meV is about 54\,\% larger
than the splitting in the system with $1\,a$ WL thickness
$\Delta^{\text{strain}}_{e_2,e_3}=7.1$ meV. Also the energy
splitting $\Delta_{e_1,e_2}$ between the ground state $e_1$ and the
first excited state $e_2$ is strongly influenced by the WL
thickness, namely $\Delta_{e_1,e_2} = 162.8$ meV for the unstrained
sytem with \mbox{$1a$ WL} but $\Delta_{e_1,e_2}=204.1$ meV for a 1
ML WL; with strain effects the splitting $\Delta_{e_1,e_2}$ is
increased by about 27\,\% if the WL thickness is decreased from
$1\,a$ to 1 ML. The results are summerized in
Table~\ref{tab:splittings}. This effect mainly arises from the fact,
that the bound states inside the dot are also coupled to the
WL-states. For a $1\,a$ WL the wave functions of the bound states
show also a probability density inside the WL. For a thinner WL the
leaking of the states into the region of the WL is much less
pronounced. In this case, the microscopic structure inside the EQD
and also the strain-affect are much more important. This explains
the larger energy splittings in case of the 1 ML WL compared to the
results for a \mbox{$1\,a$ WL}. The hole states are influenced in a
similar manner. In case of a 1 ML WL the energy spectrum of the hole
states is shifted to higher energies due to the strain-effects. This
behavior is similar to the behavior obtained from the calculations
for a $1\,a$ WL (Fig.~\ref{fig:vergl1ML}). In the 1 ML WL system the
energy splittings $\Delta_{h_1,h_2}$ and $\Delta_{h_2,h_3}$ for the
first three hole-states are larger than the values we obtain for the
system with $1\,a$ WL. These splittings are also summarized in
Table~\ref{tab:splittings}. The WL thickness also influences the EQD
energy gap $E^{\text{QD}}_\text{gap}$. For a 1 ML WL the
electron-states are shifted to higher energies in contrast to the
behavior of the hole states (compare Figs.~\ref{fig:verglbound}
and~\ref{fig:vergl1ML}). In case of the 1 ML WL the gap energy
$E^{\text{QD}}_\text{gap}$ is only slightly affected by the strain.
We observe here that the strain has opposite effect for electrons
and hole states: electron states become shallower, approaching the
conduction-band edge, while the hole states become deeper, moving
away from the valence-band edge.\\
The knowledge of the single-particle wave functions makes the
examination of many-particle effects in EQDs possible. The single
particle wave functions can be used for the calculation of Coulomb-
and dipole- matrix elements as input parameters. For example the
investigation of multi-exciton emission spectra~\cite{baer2004},
carrier capture and relaxation in semiconductor quantum dot
lasers~\cite{nielsen2004} or a quantum kinetic description of
carrier-phonon interactions~\cite{seebeck2005} is possible.
\begin{table}[t]
\begin{center}
\begin{tabular}{l|cccc}
\hline\hline\noalign{\smallskip} WL & \multicolumn{2}{c|}{1a} &
\multicolumn{2}{c|}{1 ML} \\\hline
  & No Strain  & Strain & No Strain & Strain \\\hline
$\Delta_{e_1,e_2}$ [meV] & 162.8 & 161.5 & 204.1  & 221.2 \\
$\Delta_{e_2,e_3}$ [meV] & 0.43 & 7.1 &0.5 & 10.9 \\
$\Delta_{h_1,h_2}$ [meV] & 5.76 & 3.7 & 7.25 & 7.66\\
$\Delta_{h_2,h_3}$ [meV] & 16.36 & 12.3 & 19.66 &
15.11\\\noalign{\smallskip}\hline
$E^{\text{QD}}_\text{gap}$ [eV] &  2.12 & 2.09 & 2.21 & 2.21 \\
\noalign{\smallskip}\hline\hline
\end{tabular}
\caption{Energy splittings for electron and hole bound-states in
case of different WL thicknesses. The influence of strain-effects on
the splittings $\Delta_{e_1,e_2}=|E_{e_1}-E_{e_2}|$,
$\Delta_{e_2,e_3}=|E_{e_2}-E_{e_3}|$,
 $\Delta_{h_1,h_2}=|E_{h_1}-E_{h_2}|$ and $\Delta_{h_2,h_3}=|E_{h_2}-E_{h_3}|$ is also displayed. The WL thickness is $1\,a$
and one monolayer (1ML), respectively. The base length of the
pyramid is $b=10\,a$.} \label{tab:splittings}
\end{center}
\end{table}

\section{Results for $\bf CdSe$ Nanocrystals}
\label{sec:resultsnano}
\subsection{Geometry and Strain}

In this section we investigate the single particle states of CdSe
nanocrystals within our TB-model. These
nano\-structures are chemically synthesized~\cite{kim2001,guzelian96}.
The nanocrystals are nearly spherical in
shape~\cite{banin,guzelian96,alivisatos96}
and the surface is passivated by organic ligands.
Due to the flexible surrounding matrix, these nanostructures are
nearly unstrained~\cite{alivisatos96}.
The size of these nanostructrues is in between 10 and 40 \AA\ in
radius~\cite{banin98,banin,alperson99,kim2001}.

We model such a chemically synthesized NC as an unstrained,
spherical crystallite with perfect surface passivation. The
zincblende structure is assumed for the CdSe nanocrystal. We neglect
surface reconstructions~\cite{pokrant99,puzder2004-2,deglmann2002}
and that the surface coverage with ligands is often not
perfect~\cite{taylor2002}, though these effects can be important
especially for very small NCs. However, we concentrate on
considerably larger NCs than in the before mentioned referenceses.
Therefore, unlike previous TB work we concentrate here on size and
the size dependence of the results obtained for the electronic
structure of the NCs. The TB-parameters, which describe the coupling
between the dot material and the ligand molecules, are chosen to be
zero. This corresponds to an infinite potential barrier at the
surface and is commonly used because of the larger band gap of the
surrounding material~\cite{lippens89}. An alternative approach to
treat the ligand molecules is discussed by Sapra \textit{et al.} in
Ref.~\onlinecite{sapra2004}. The influence of the organic ligands on
the electronic structure can also be investigated more realistically
in the framework of microscopic
descriptions~\cite{pokrant99,rabani2001,puzder2004-2}.

\subsection{Single particle states and comparsion with experimental results}

\begin{figure*}
\centering
\begin{tabular}{c@{\qquad}c@{\qquad}c@{\qquad}c@{\qquad}c}
\multicolumn{5}{c}{\textbf{Electron and holes states for the nanocrystal}}\\
\hline\hline
\multicolumn{5}{c}{Electron states}\\
\hline
$e_1$ & $e_2$ & $e_3$ & $e_4$ & $e_5$ \\
  \resizebox{0.15\textwidth}{!}{\includegraphics{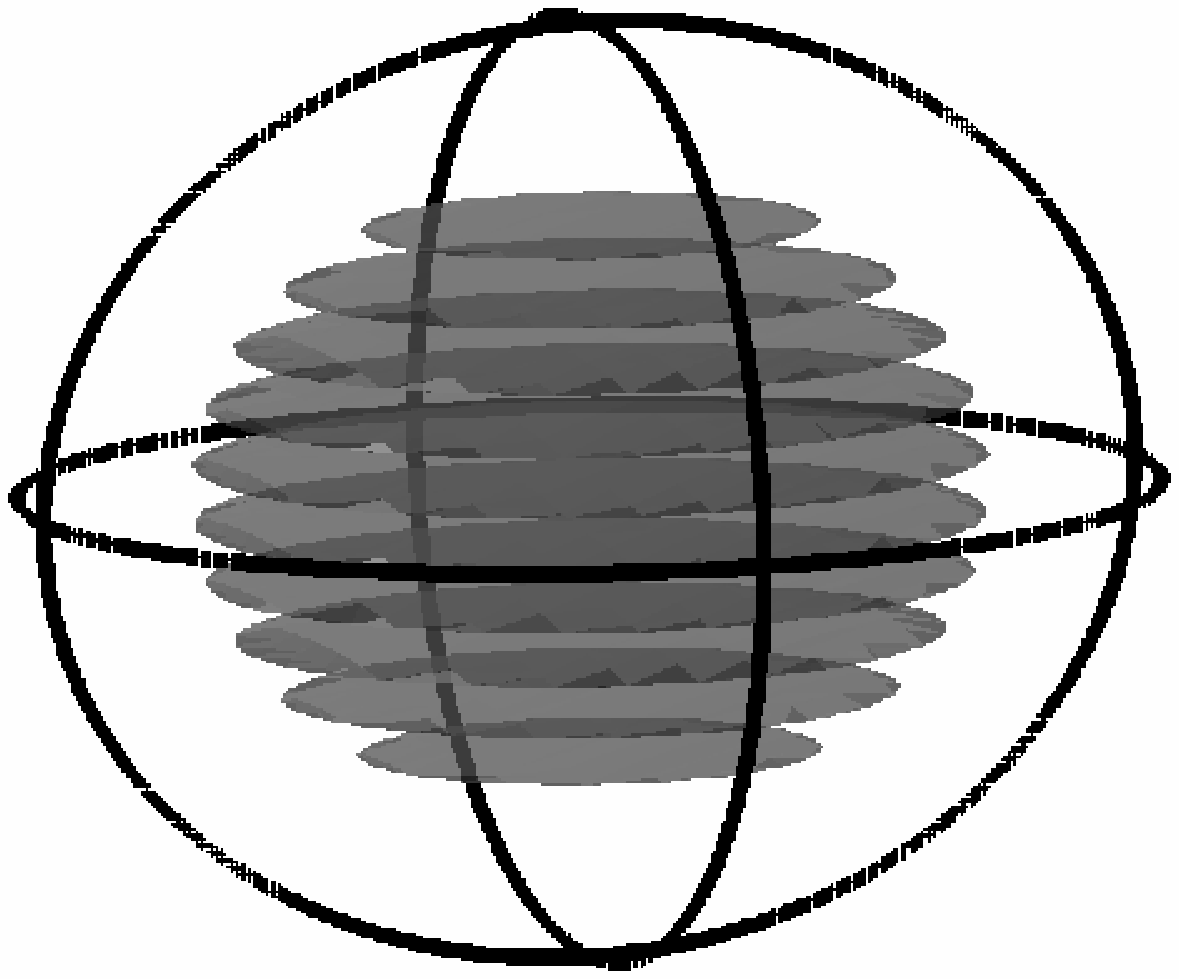}} &
  \resizebox{0.15\textwidth}{!}{\includegraphics{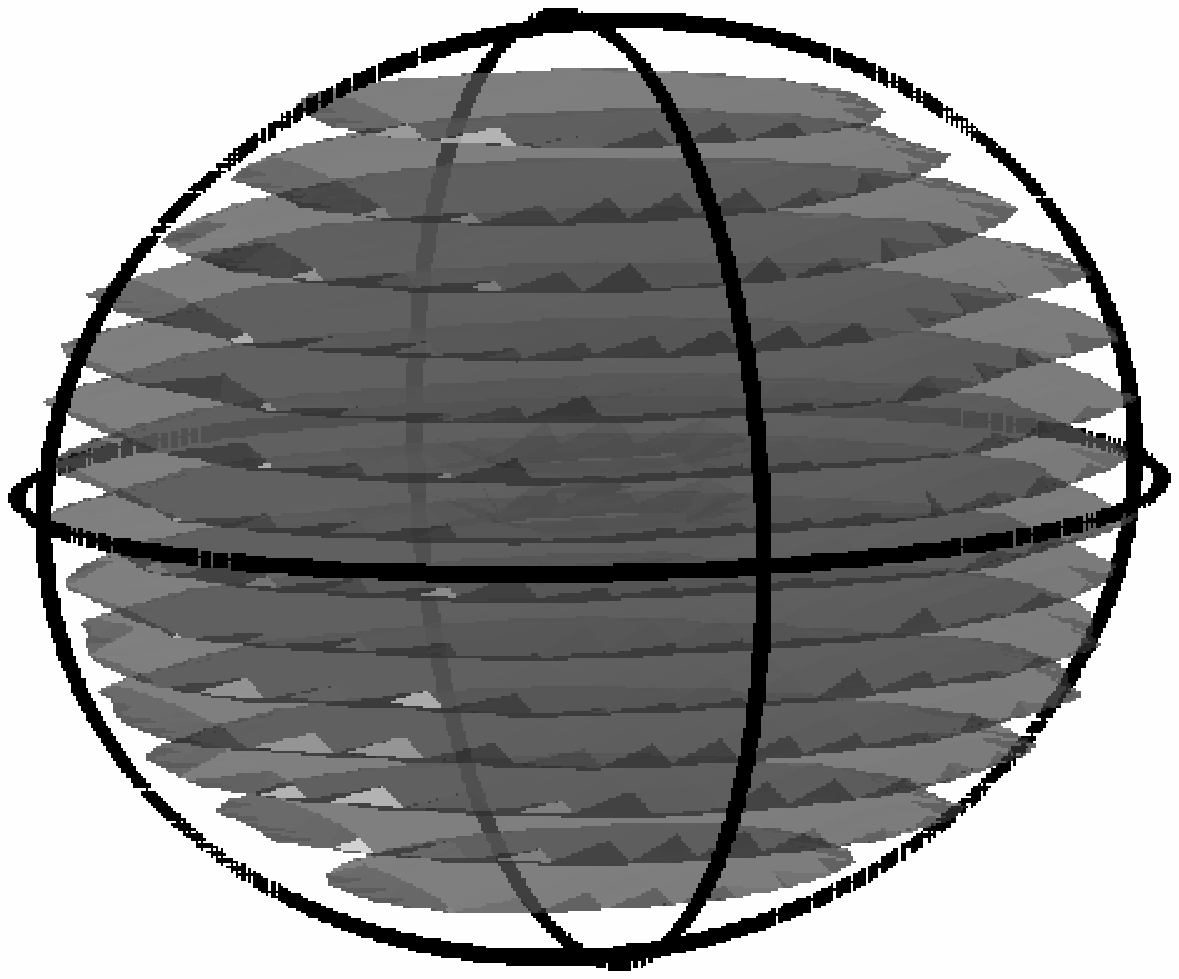}} &
  \resizebox{0.15\textwidth}{!}{\includegraphics{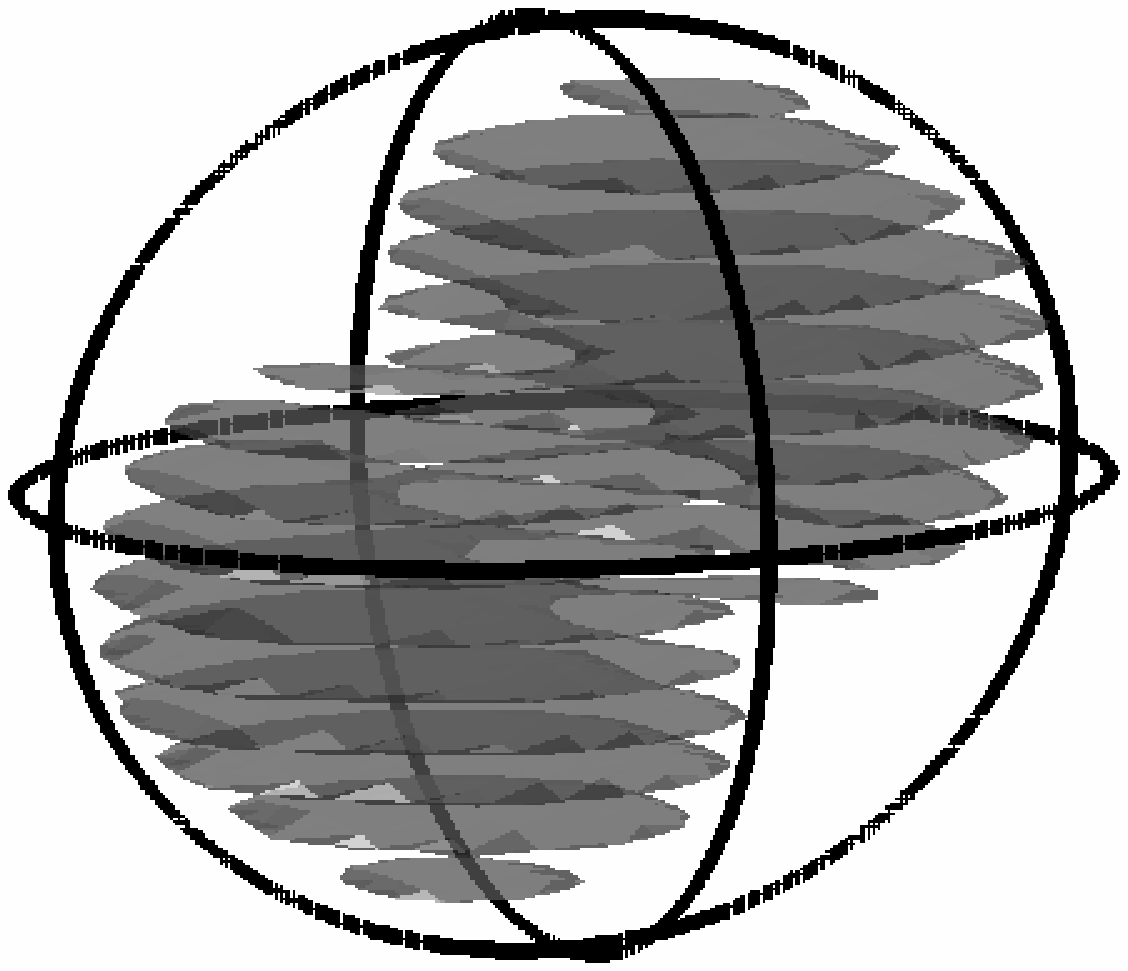}} &
  \resizebox{0.15\textwidth}{!}{\includegraphics{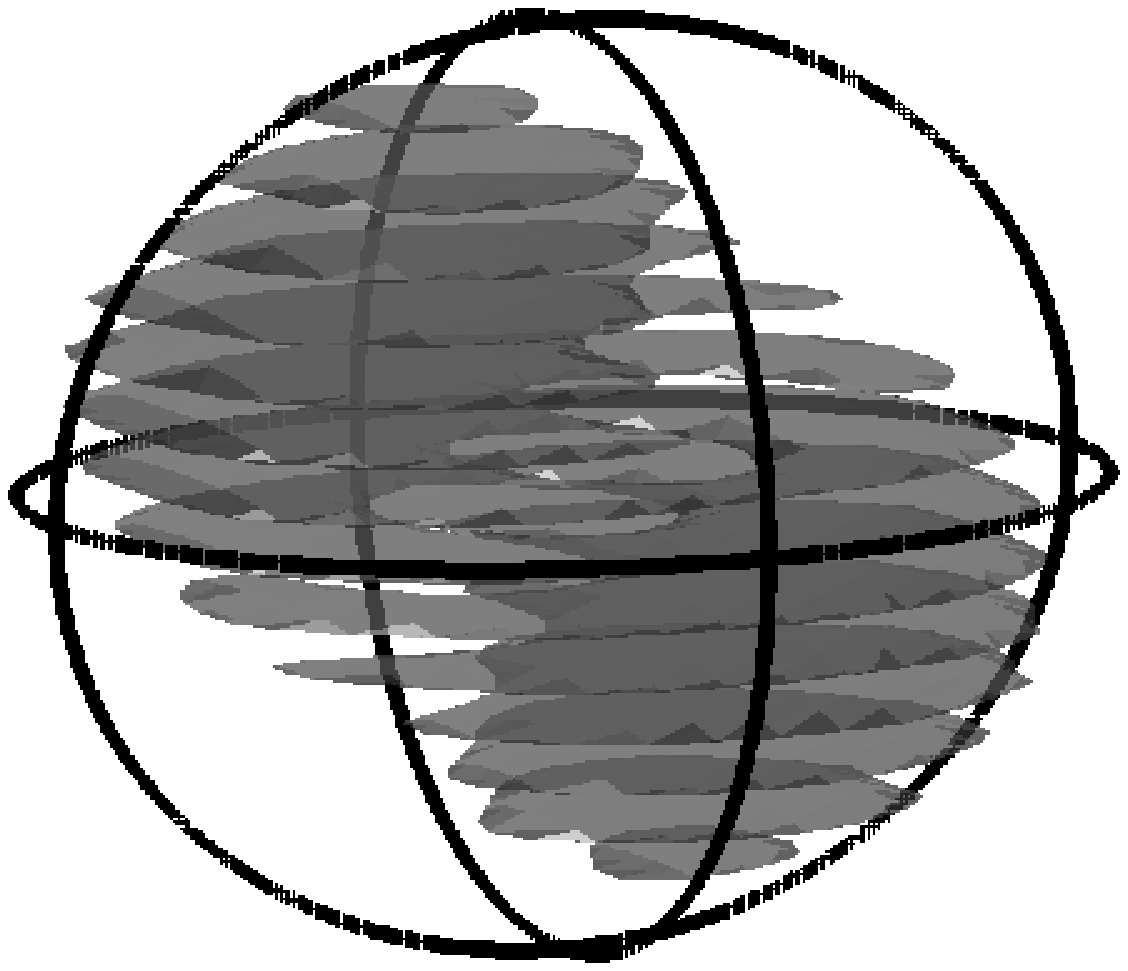}} &
  \resizebox{0.15\textwidth}{!}{\includegraphics{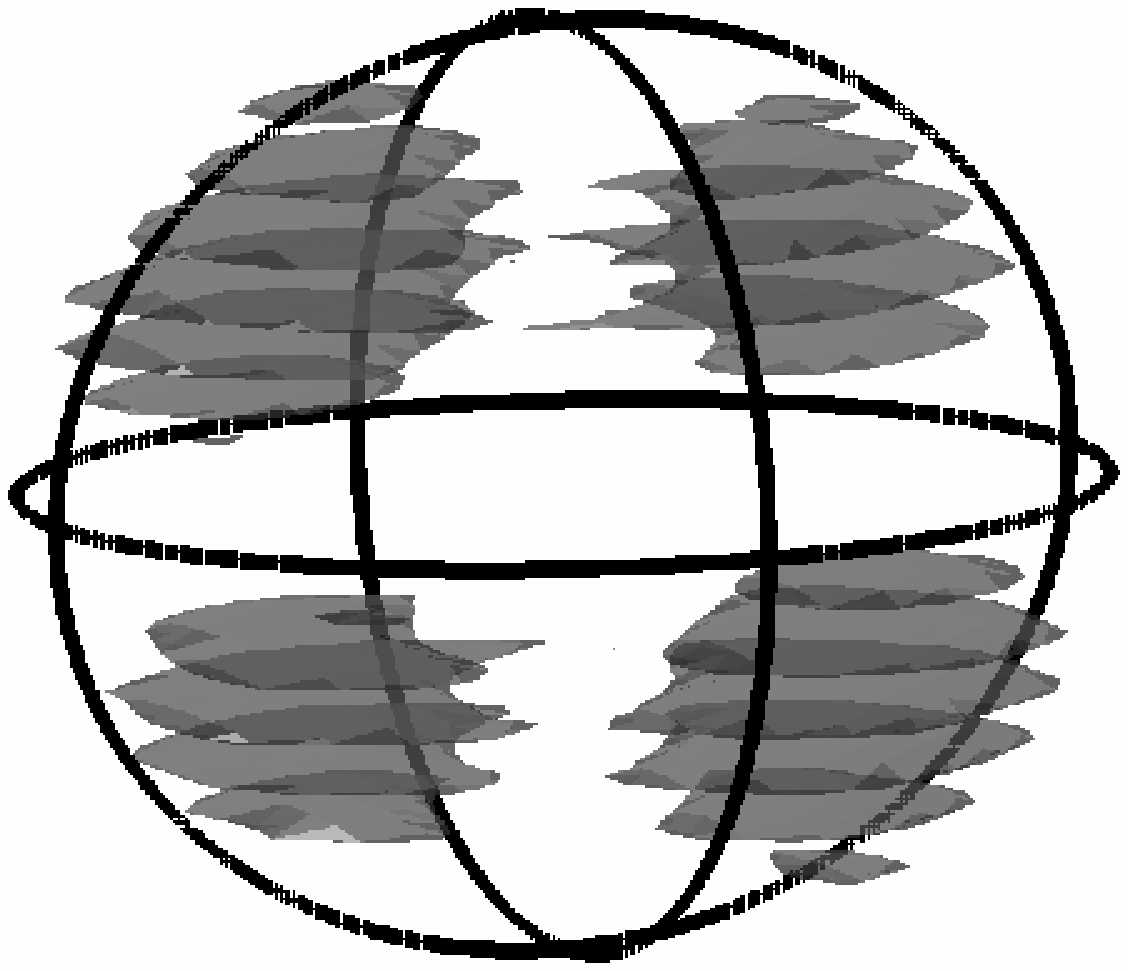}} \\
  \\
\multicolumn{5}{c}{Hole states}\\
\hline
$h_1$ & $h_2$ & $h_3$ & $h_4$ & $h_5$ \\
  \resizebox{0.15\textwidth}{!}{\includegraphics{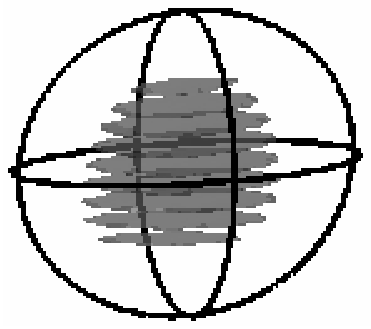}} &
  \resizebox{0.15\textwidth}{!}{\includegraphics{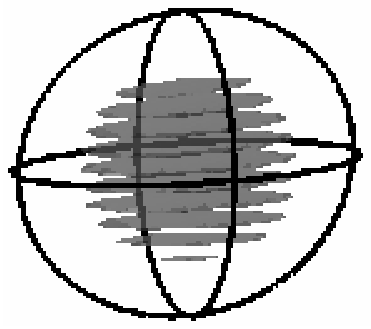}} &
  \resizebox{0.15\textwidth}{!}{\includegraphics{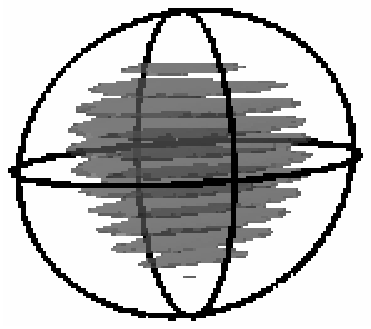}} &
  \resizebox{0.15\textwidth}{!}{\includegraphics{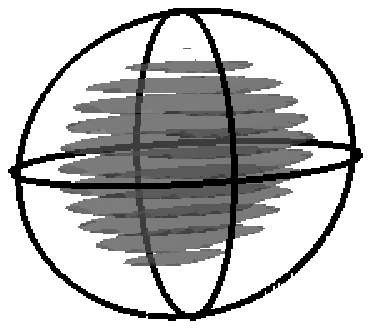}} &
  \resizebox{0.15\textwidth}{!}{\includegraphics{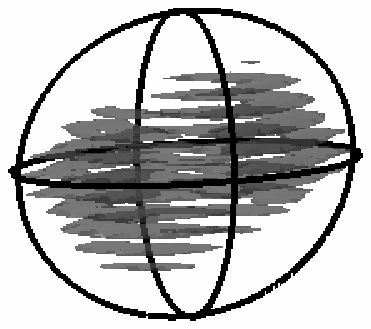}} \\
  \\
  \hline\hline
\end{tabular}
\caption{Isosurfaces (at 30 \% of the maximum probability density) of the squared electron and hole wavefunctions of
spherical CdSe nanocrystals of diameter $d=4.85$ nm for the five lowest states}
\label{fig:electrholeNCstates}       
\end{figure*}

We have performed TB-calculations for finite, spherical, unstrained
NCs of diameter between 1.82 nm and 4.85 nm (corresponding to 3-8
$a$, when $a \approx 6.07$ {\AA} is the CdSe lattice constant of the
conventional unit cell). The finite matrix diagonalizations yield
both, the discrete eigenenergies and the eigenstates. For the
largest NCs (of diameter 4.85 nm) results for the five lowest lying
electron and hole eigenstates are shown in Fig.
\ref{fig:electrholeNCstates} again in the form of an isosurface
plot. The lowest lying electronic state $e_1$ obviously has
spherical symmetry and can be classified as a $1s$-state.
Correspondingly the second state $e_2$ has the form of a $2s$-state
and the states $e_{3,4}$ are $p$-states and $e_5$ is a $d$-like
state. Despite the spherical symmetry of the system this simple
classification is no longer possible for the hole states, however.
Even the lowest lying hole state $h_1$ has no full rotational
invariance, i.e. strictly speaking it cannot be classified as being
an $s$-state. This is due to the intermixing of different atomic
TB-valence electron states in the NC. Similarly the higher hole
states $h_2 - h_4$ cannot clearly be classified as an $s$- or
$p$-like state. This is an effect, which simple effective mass
models cannot account for, but which will have implications in the
calculation of matrix elements between these states, which enter
selection rules for optical transitions etc.

In the case of an ideal zinc blende structure as considered here we
do not obtain any indications of quasi-metallic behavior, i.e. of a
non-vanishing (quasi continuous) spectrum of states  at the Fermi
energy in contrast to previous work (assuming an ideal wurtzite
structure for CdSe
nanocrystals)~\cite{puzder2004,leung99,pokrant99}. This is probably
due to the fact that this quasi-metallic behavior is due to surface
states in the case when no passivation and surface reconstruction is
taken into account. These surface states are formed by the dangling
bonds  of unsaturated Se at the NC surface, which cause $s$-states
in the band gap region~\cite{leung99}. In our simplified and
restricted TB $s_cp^3_a$-basis set these $s$-orbitals at the anions
(Se) are not taken into account. Therefore, these surface states,
which in reality and in more realistic models are removed (i.e.
energetically drawn down and filled) due to passivation and surface
reconstruction, do not occur.

The discrete electronic states of semiconductor NCs are
experimentally accessible by scanning tunneling microscopy
(STM)~\cite{banin,alperson99,banin98}. The tunnel current $I$
between the metallic tip of the STM and the CdSe nanocrystal, which
is e.g. epitaxially electrodeposited onto a template-stripped gold
film, is measured as a function of the bias voltage $V$. The
conductance ($dI/dV$) is related to the local tunneling density of
states. In the $dI/dV$ versus $V$ diagram, several discrete peaks
can be observed. These peaks  correspond to the addition energies
(charging energies) of holes and electrons. The spacing between the
various peaks can be attributed to the Coulomb charging (addition
spectrum) and/or charge transfer into higher energy levels
(excitation spectrum). From these measurements the energy gap
$E^{\text{nano}}_{\text{gap}}$ as well as the splitting
$\Delta_{e_1,e_2}$ between electron ground state $e_1$ and the first
excited state $e_2$ can be determined.

\begin{figure}[]
\begin{center}
\resizebox{0.4\textwidth}{!}{%
  \includegraphics{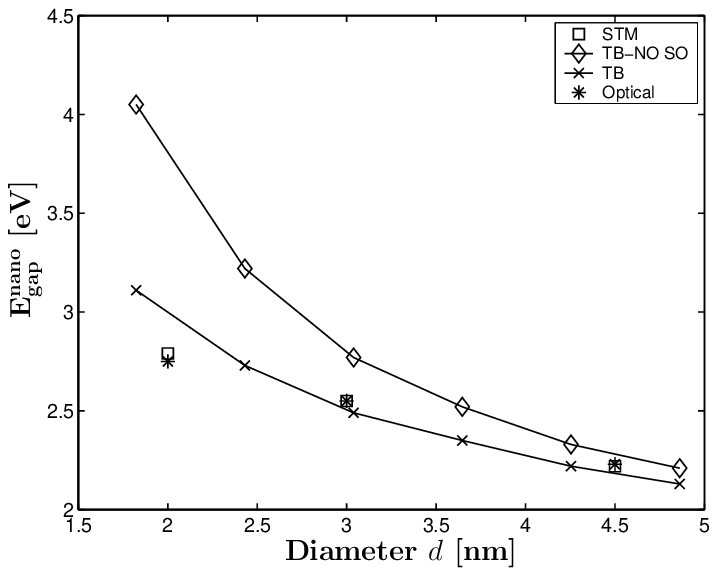}
}
\caption{Energy gap $E^{\text{nano}}_{\text{gap}}$ as a function of the nanocrystal diameter $d$. Compared are the
results from our TB-model with (TB) and without (TB-NO SO) spin-orbit coupling, a STM
(STM)~\cite{alperson99} and an optical measurement (Optical)~\cite{alperson99}.}
\label{fig:nanogap}       
\end{center}
\end{figure}

\begin{figure}
\begin{center}
\resizebox{0.45\textwidth}{!}{%
  \includegraphics{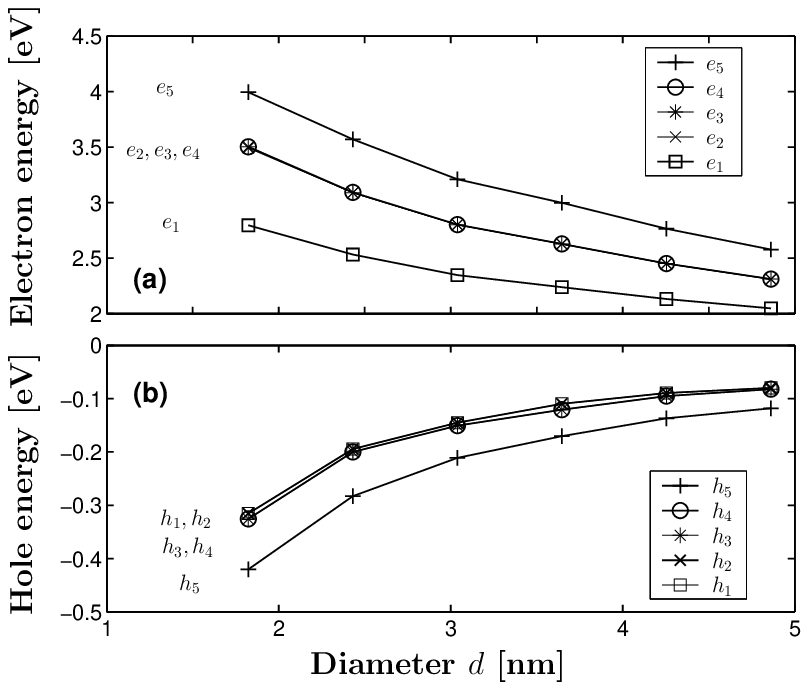}
}
\caption{Electron (a) and hole (b) energies as a function of the nanocrystal diameter $d$. For electrons ($e_1$-$e_5$)
and holes ($h_1$-$h_5$) the first five eigenvalues are displayed. Each state is twofold degenerated.}
\label{fig:eigennano}       
\end{center}
\end{figure}

Alperson \textit{et al.}~\cite{alperson99} investigated CdSe
nanocrystals with an STM. Here  we compare our calculated energy gap
$E^{\text{nano}}_{\text{gap}}$, which is given by the difference
between the electron, $e_1$, and hole, $h_1$, groundstate, with
measured data from Ref.~\onlinecite{alperson99}.
Figure~\ref{fig:nanogap} displays the results for CdSe NCs with
diameters in between $1.82$ nm and $4.85$ nm. Alperson~\textit{et
al.}~\cite{alperson99} compare the STM results (dashed dotted line)
with optical spectroscopy measurements (dotted line) from Ekimov
\text{et al.}~\cite{ekimov93}. The overall agreement with the TB
results is very good, especially for the larger NCs. Deviations in
the case of the small 2 nm NC arise from surface
reconstructions~\cite{puzder2004,deglmann2002,pokrant99} which are
neglected here. When the same calculation is done without spin-orbit
coupling (TB-NO SO), the energy gap $E^{\text{nano}}_{\text{gap}}$
is always strongly overestimated by the TB-model, in particular for
smaller nanocrystals. So the spin-orbit coupling is important for a
satisfactory reproduction of the experimental results. For the
calculations without spin-orbit coupling, the TB-parameters are
re-optimized to reproduce the characteristic properties (band gap,
effective masses) of the bulk material. The re-optimized parameters
are given in Table \ref{tab:TB-parameters}.
Taking into account the electron spin, the lowest electron state $e_1$ is twofold degenerated and \mbox{$s$-like}. This
is consistent with the experimentally observed doublet~\cite{alperson99} in the $dI/dV$ characteristic.
The next excited level is (quasi) sixfold degenerated. The spin-orbit coupling splits
this into one twofold and one fourfold degenerate state~\cite{niquet2001}.
In the STM measurement Alperson \textit{et al.}~\cite{alperson99} observed such a higher multiplicity
of the second group of peaks.
This behavior has also experimentally~\cite{banin} and theoretically~\cite{niquet2001}
been observed for
InAs nanocrystals. The electron energy spectrum
for NCs of different diameter is shown in Fig.~\ref{fig:eigennano} (a). Here the first five
electron states $e_1-e_5$ are displayed. Note that every state is twofold degenerated due to the spin.\\
For the hole states the situation is more complicated.
Alperson~\textit{et al.}~\cite{alperson99} observed a high density
of states at negative bias. The distinction between addition and
excitation peaks is difficult, due to the large number of
possibilities and the close proximity between the charging energy
and the level spacing. For the holes we obtain that the first two
states $(h_1,h_2)$ and $(h_3,h_4)$ are fourfold degenerated. The
energy splitting of these states is also very small. These results
are consistent to the observations of Alperson~\textit{et
al.}~\cite{alperson99}. Figure~\ref{fig:eigennano} (b) shows the
hole energy versus diamater $d$ for the spherical CdSe NCs.
Obviously, for all diameters displayed the states $h_1 - h_4$ are
almost degenerate, i.e. including spin there is almost an 8-fold
degeneracy
of these states.\\
\begin{figure}[]
\begin{center}
\resizebox{0.4\textwidth}{!}{%
  \includegraphics{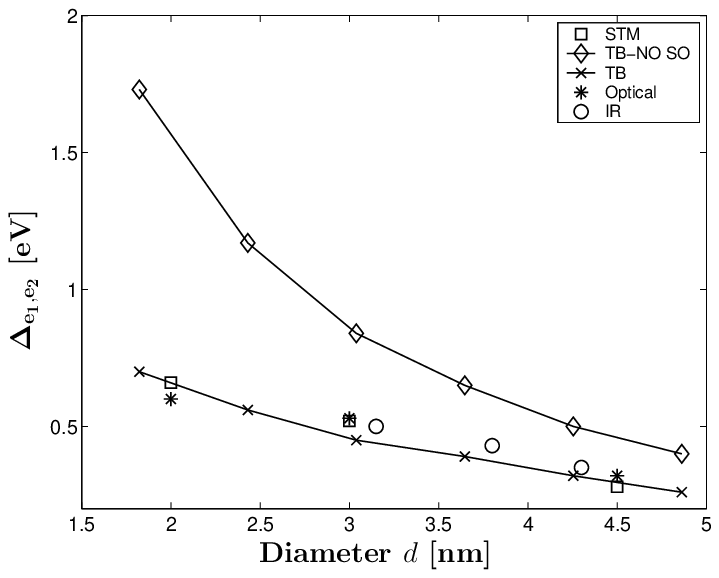}
}
\caption{Splitting $\Delta_{e_1,e_2}=E_{e_2}-E_{e_1}$ between the lowest two
electronic states as a function of the nanocrystal diameter $d$.
The results from our TB-model, with (TB) and without (TB-NO SO) spin-orbit coupling, and from
an STM measurement (STM, Ref.~\onlinecite{alperson99}) are
displayed. Besides this results from infrared spectroscopy (IR, Ref.~\onlinecite{guyot98}) and optical methods
(Optical, Ref.~\onlinecite{alperson99}) are shown.}
\label{fig:splittingnano}       
\end{center}
\end{figure}
Furthermore the calculated splitting $\Delta_{e_1,e_2}=E_{e_2}-E_{e_1}$
between the first two electron
states $e_1$ and $e_2$ is compared with experimentally observed results for this quantity.
Figure~\ref{fig:splittingnano} shows
$\Delta_{e_1,e_2}$ as a function of the nanocrystal diameter $d$.
The influence of the
spin-orbit coupling on our results is also investigated.
We have done the calculations without (TB-NO SO) and with spin orbit-coupling (TB).
The results of our TB-model for the splitting $\Delta_{e_1,e_2}$ are
compared with results obtained by
STM~\cite{alperson99} and by optical methods (optical)~\cite{ekimov93}.
This splitting $\Delta_{e_1,e_2}$ was independently determined experimentally by Guyot-Sionnest and Hines~\cite{guyot98} using
infrared spectroscopy (IR).
Without spin-orbit coupling the TB-model always overestimates the splitting $\Delta_{e_1,e_2}$.
Especially for smaller nanocrystals the spin-orbit coupling is very important to describe the electronic structure.
With spin-orbit coupling the results of the TB-model show good agreement with the experimentally observed results.

\section{Conclusion}
\label{sec:conclusion}

We have applied an empirical $s_cp_a^3$ TB-model to the calculation
of the electronic properties of II-VI semiconductor EQDs and NCs.
Assuming a zinc blende lattice and (per spin direction) one $s$-like
orbital at the cation sites and three $p$-orbitals at the anion
sites, the TB-parameters for different materials (here ZnSe and
CdSe) are determined so that the most essential properties (band
gap, effective masses etc.) of the known band structure of the (3
dimensional) bulk materials are well reproduced by the TB band
structure. Then a CdSe QD (on top of a two-dimensional, a few atomic
layers thick WL) embedded within a ZnSe matrix is modelled by using
the TB-parameters of the dot material for those sites occupied by
CdSe and the ZnSe TB matrix elements for the remaining sites;
suitable averages have to be chosen for intersite matrix elements
over and for on-site matrix elements on anion (Se) sites at
interfaces between QD and barrier material. Spherical CdSe NCs can
be modelled similarly by setting the intersite matrix elements
between surface atoms and atoms in the monolayer of surfactant
material to zero. The effects of the spin-orbit interaction, the
band offsets and for the EQDs strain effects are taken into account.

For the EQD systems the numerical diagonalization yields a discrete
spectrum of bound electron and hole states localized in the region
of the EQD. Energetically these discrete states are below the
continuum of the WL states. We have investigated the dependence on
the EQD size and find that the number of the bound states and their
binding energy increases with increasing dot size, therefore the
effective band gap decreases. We have also investigated the
dependence of the bound eigenenergies and their degeneracy on strain
and on the thickness of the WL. Looking at the states themselves one
sees that conduction band (electron) states can be roughly
classified as $s$-like, $p$-like, etc.  states but the valence band
(hole) states cannot be classified according to such simple
($s$,$p$,$d$) symmetries because they are determined by a mixing
between the different (anion) $p$-states. This cannot be accounted
for by simple effective mass models but it will be important for
instance for the calculation of dipole matrix elements between
electron and hole states which determine the selection rules for
optical transitions. For the NCs the whole spectrum is discrete, but
in spite of the spherical symmetry the hole states do not have the
simple $s$,$p$,$d$-symmetry but are intermixtures of atomic
$p$-orbitals. Even the lowest hole state has no spherical
$s$-symmetry but it is 4-fold (8-fold including spin) degenerate.
The spin-orbit interaction is very important. Including the
spin-orbit interaction we obtained nearly perfect agreement with
experimental results obtained by STM for the dependence of the band
gap and of the splitting of the lowest electronic states on the
diameter of the NC.

Compared to (two-band) effective mass~\cite{wojs96,grundmann1} and
multi-band
$\vec{k}\cdot\vec{p}$-models~\cite{grundmann98,fonoberov2003,cpryor98}
for EQDs our TB model clearly has the advantage of a microscopic,
atomistic description. Different atoms and constituents of the
nanostructure and their actual positions are considered, and this
may lead to a reduction of symmetries (for instance the $C_{2v}$
symmetry instead of a $C_{4v}$-symmetry). This may automatically
lift certain degeneracies and lead automatically to a splitting, for
instance between $e_2$ and $e_3$ states, whereas an
8-Band-$\vec{k}\cdot \vec{p}$ model still yields degenerate $e_2$
and $e_3$ states~\cite{cpryor98}. The effects of inhomogeneous
strain can be easily incorporated into a TB model by considering the
deviations of the actual atomic positions from the ideal position in
the bulk crystal. Only the (empirical) pseudopotential
treatment\cite{wang99,wang2000,williamson2000,kim98} may be still
superior and more accurate than the TB approach, but in a
pseudopotential descripion a variation of the wave functions within
the individual atoms is accounted for and a large number of basis
states is required. Therefore, a TB description is simpler and
quicker and allows for the investigation of larger nanostructures
without loosing information on the essential, microscopic details of
the structure. Compared to other TB models of QD structures, we do
not consider free standing, isolated QDs (as in
Ref.~\onlinecite{saito98}) but we can describe realistic QDs (with a
WL) embedded into another barrier material. We show here that a
reduced $s_cp_a^3$-basis is already sufficient for a satisfying
reproduction of properties like optical gaps, energy splittings,
etc., and their size dependence. Much larger basis sets, namely a
$sp^3s^*$-basis\cite{santo}  or even a
$sp^3d^5s^*$-basis\cite{jancu98} were used in previous TB-models of
EQDs. Our reduced, smaller basis set, of course, leads to
computational simplifications and allows for the treatment of larger
QDs. Furthermore, we apply our TB-model to different materials than
investigated previously, namely II-VI CdSe nanostructures, and we
investigate also NCs, for which excellent agreement with
experimental STM-results could be demonstrated.

In the future further applications of our TB-model for embedded
semiconductor QDs and NCs are planned. Of course applications to QDs
of other materials, for instance nitride systems, and other (e.g.
wurtzite) crystal structures are possible. Furthermore, EQDs of
other shape and size (dome-shaped, lens-shaped, truncated cones,
etc.) or two coupled QDs or freestanding (capped and uncapped) QDs
can be investigated. A combination with ab-initio calculations is
also possible by determining the TB-parameters from a
first-principles band structure calculation of the bulk material.
Furthermore the influence of surface reconstructions and the
surfactant material on the results for NCs should be investigated.
Especially for small NCs these effects are important. Finally matrix
elements of certain observables like dipole matrix elements between
the calculated QD electron and hole states can be determined, which
are important for selection rules and the optical properties of
these systems.

\begin{acknowledgments}
   This work has been supported by a grant from the Deutsche
   Forschungsgemeinschaft No. Cz-31/14-1,2 and by a grant
for CPU time from the John von Neumann Institute
   for Computing at the Forschungszentrum J\"ulich.
\end{acknowledgments}

%
%
 \bibliographystyle{apsrev}
 \bibliography{literaturges}

\end{document}